\documentclass[12pt]{article}
\usepackage{amsmath,amscd,mathrsfs,bm,amssymb,color,theorem,hiroshima}
\usepackage{latexsym}
\newtheorem{theorem}{Theorem}[section]
\newtheorem{proposition}[theorem]{Proposition}
\newtheorem{lemma}[theorem]{Lemma}
\newtheorem{corollary}[theorem]{Corollary}

\newtheorem{example}[theorem]{Example}
\newtheorem{remark}[theorem]{Remark}

\makeatletter
\@addtoreset{equation}{section}
\makeatother

\title
{\sc Fiber decomposition of non-commutative harmonic oscillators 
 by \\ two-photon quantum Rabi models}
\author{Fumio Hiroshima\footnote{corresponding author: hiroshima@math.kyushu-u.ac.jp}\\ Faculty of Mathematics, Kyushu university, \\ 
Fukuoka, 819-0395, Japan\\ 
and \\
Tomoyuki Shirai\\
Institute of Mathematics for Industry, Kyushu university, \\ 
Fukuoka, 819-0395, Japan}
\date{\today}
\begin{document}
\maketitle
\begin{abstract}
The non-commutative harmonic oscillators $\nec$ with parameters $\alpha,\beta>0$, $\alpha\beta>1$ and 
two-photon quantum Rabi models $\rab$ with $\triangle\geq 0$, $|g|<\frac{1}{2}$, are 
both extensions of the one-dimensional harmonic oscillator.
In the special case where $\alpha=\beta$ and $\triangle=0$ it is immediately seen that
$Q_{\alpha\alpha}^{\rm NH}$ is unitarily equivalen to $
\frac{\sqrt{\alpha^2-1}}{\sqrt{1-4g^2}}H_{0,g}^{\rm 2p}$.
The purpose of this paper 
is to establish relationships between 
$\nec$ and $\rab$ for the general cases $\triangle>0$ and $\alpha\neq\beta$, 
and to show the fiber decomposition 
of $\nec$
in terms of $\rab$. 
We also construct 
Feynman-Kac formulas 
for $e^{-t\rab}$ and $e^{-t \nec} $. 
It is then considered 
the asymptotic behaviors of the spectral zeta function 
$\zeta_{\rm 2p}(s)$ of $\rab$. 
\end{abstract}

{

\section{Introduction}
In this paper we are concerned with relationships between  
non-commutative harmonic oscillators (NcHO) and 
two-photon quantum Rabi models  from an 
operator-theoretic point of view, and study the asymptotic behaviors of the spectral zeta function of 
the two-photon quantum Rabi models using path measures. 
For brevity, we refer to the two-photon quantum Rabi model as 2pQRM.

The NcHO is a second order differential operator 
and had been introduced by Parmeggiani and Wakayama \cite{PW01,PW02a,PW02b} as a non-commutative extension of the one-dimensional harmonic oscillator through 
the oscillator representation of Lie algebra ${\mathfrak sl}_2(\RR)$, and 
the spectrum of the NcHO are studied so far from a purely mathematical point of view. 
In the Bargmann representation the eigenvalue equation of the NcHO 
can be reduced to studying Heun ordinary differential equations \cite{och01}. 
While the spectrum of the NcHO is purely discrete and one can define its spectral zeta function. 
A meromorphic continuation and special values of the spectral zeta function 
have been investigated in \cite{IW05a,IW05b,KW06,IW07,och08,rey23,KW23}.

On the other hand the standard quantum Rabi model describes a two-level atom coupled with a single photon. 
In this paper the standard quantum Rabi model is called one-photon quantum Rabi model and we abbreviate as  1pQRM. 
It is shown in \cite{bra11} that the 1pQRM is integrable due to the parity symmetry, 
and as a result, the spectrum of the 1pQRM has been extensively studied. 
There are many generalizations of the 1pQRM which describe various implementations within cavity and circuit QED as well as quantum simulation platforms. 
See \cite{bra23} and references therein. 
The so-called two-photon quantum Rabi model, 2pQRM, 
is one generalization, which is the main subject of the present paper. 
The 2pQRM is also a second order differential operator 
and the spectrum of the 2pQRM are studied so far but mainly from a physical point of view. 
The amount of research on the 2pQRM is vast, and it is difficult to cover it all here. 
For example, the spectrum of the 2pQRM is studied in \cite{DXBC16}, 
the kpQRM ($k\geq3$) is discussed in \cite{bra24}, 
the hybrid of the 2pQRM and the 1pQRM is introduced in \cite{FLH19}, 
the degeneracy of the spectral curve is investigated in \cite{FH21}, 
and the case of critical coupling $g=\half$ 
is studied in \cite{AMR20}.

Although the NcHO and the 2pQRM appear  quite similar, as shown below, they have often been studied independently. In particular, the 2pQRM has been valued as the simplest yet nontrivial model in physics, whereas the NcHO has 
been primarily discussed in terms of number theory and complex differential equations. 
It is remarkable, however, that \cite{nak24} was recently published as a paper bridging the~NcHO and 
the~2pQRM, where it was shown that the eigenvectors of the NcHO coincide with those of a certain 2pQRM. 
This result serves as one of the motivations for the present paper.

In this paper we discuss (1) relationships between $\nec$ and $\rab$, 
(2) Feynman-Kac formulas of $e^{-t\rab}$ and $e^{-t\nec}$,
(3) the asymptotic behaviors of the spectral zeta functions of $\rab$. 
In what follows we explain (1), (2) and (3). 

(1) We show that the NcHO can be decomposed into the 2pQRM. 
The non-commutative harmonic oscillator 
\[\nec =\AB\otimes \left( 
\add a +\half\right) +\half\J \otimes (a^2-{\add}^2)\]
 is defined by two parameters $\alpha,\beta>0$, and 
two-photon quantum Rabi model 
\[\rab =\triangle\begin{pmatrix}1&0\\0&-1\end{pmatrix}
\otimes \one+\one\otimes 
\left( \add a+\half\right)+ 
g\begin{pmatrix}0&1\\1&0\end{pmatrix}\otimes (a^2+{\add}^2)\]
by $\tri>0$ and a coupling constant $g\in\RR$. 
See Section \ref{2} for the detail. 
Both $\nec$ and $\rab$ are defined as self-adjoint operators in 
the Hilbert space:
$\cH= \CC^2\otimes\LR$. 
We can see that $\rab$ is unitarily equivalent to the form 
\begin{align}\label{r1}
\rab\cong H_{0,g}^{\rm 2p}+\tri \M{0&1\\ 1 &0},
\end{align}
where  
$H_{0,g}^{\rm 2p}$ is diagonalized and its spectrum is exactly specified.  
On the other hand $\nec$ can be represented as 
\begin{align}\label{r2}
\nec=\M{\sqrt\alpha&0\\ 0& \sqrt\beta}
\nek \M{\sqrt\alpha&0\\ 0& \sqrt\beta},
\end{align} where $\nek$ 
is also diagonalized and its spectrum is exactly specified as is $H_{0,g}^{\rm 2p}$.
Note that 
$\MM{\sqrt\alpha&0\\ 0& \sqrt\beta}$ is not unitary whenever $\alpha\neq \beta$. 
See Lemma~\ref{re} for $H_{0,g}^{\rm 2p}$ and Lemma \ref{K} for $\nek$. 
Both models have very simple forms, and when a mathematician looks at them for the first time, they may appear trivial. Their spectral analysis is however not straightforward, as demonstrated in this paper.
When $\alpha=\beta$ and $\tri=0$, 
by \eqref{r1} and \eqref{r2} it is trivial to see that 
\[Q_{\alpha\alpha}^{\rm NH}\cong 
\frac{\sqrt{\alpha^2-1}}{\sqrt{1-4g^2}}H_{0,g}^{\rm 2p}, \quad 
|\alpha|>1,\quad |g|<\half.\]
The main purpose of this paper is to give relationships 
between $\nec$ and $\rab$ but for $\alpha\neq \beta$ and $\tri>0$. 
New ingredient in this paper is to 
introduce a Hilbert space $\cHa$ for each pair $\alpha,\beta>0$ in \eqref{cha}. 
It is given by 
\[(f,g)_{\alpha\beta}=(f, \gamma_{\alpha\beta}g)_\cH,\] 
where $\gamma_{\alpha\beta}$ is the matrix defined by 
\[\gamma_{\alpha\beta}=\M{1/\alpha&0\\ 0&1/\beta}\otimes \one.\] 
Then 
for each eigenvalue $\la$ of $\nec$ 
one defines a two-photon quantum Rabi model $\ral$ with $\triangle=\frac{\alpha-\beta}{2\alpha\beta}\la$ and $g=\frac{1}{2\sqrt{\alpha\beta}}$, which acts in $\cHa$, and it can be shown that 
\begin{align}
\label{31}
\rI_{\alpha\beta}
\nec 
\rI_{\alpha\beta}
^{-1}
=\frac{2\alpha\beta}{\alpha+\beta}\bigoplus_{\la\in \sss(\nec)} 
\ral \rP_{\frac{\alpha+\beta}{2\alpha\beta}\la}
\end{align}
holds on $\cHa$ in Theorem~\ref{main1}. 
Here $\rI_{\alpha\beta}$ is a unitary operator from $\cH$ to $\cHa$ and 
$\rP_{\frac{\alpha+\beta}{2\alpha\beta}\la}$ 
denotes the projection onto the eigenspace of 
$\ral$ corresponding to  eigenvalue $\frac{\alpha+\beta}{2\alpha\beta}\la$. 
In this paper we also introduce 
one-particle non-commutative harmonic oscillator 
\[\necc=\AB\otimes\lk \add a+\half\rk+\half \J (a-\add)\]
and establish relationships between $\necc$ and 
one-photon quantum Rabi model
\[\rabb=\triangle\begin{pmatrix}1&0\\0&-1\end{pmatrix}\otimes \one+\one\otimes 
\lk \add a+\half\rk+ g \begin{pmatrix}0&1\\1&0\end{pmatrix} \otimes (a+{\add})\]
 in 
Theorem~\ref{main4}. The result is 
\begin{align}
\label{41}
\rI_{\alpha\beta}
\necc \rI_{\alpha\beta}
^{-1}
=
\frac{2\alpha\beta}{\alpha+\beta}\bigoplus_{\la\in \sss(\necc)} 
\rall 
\tilde \rP_{\frac{\alpha+\beta}{2\alpha\beta}\la}, 
\end{align}
where $\tilde \rP_{\frac{\alpha+\beta}{2\alpha\beta}\la}$ 
denotes the projection onto the eigenspace of 
$\rall$ corresponding to eigenvalue $\frac{\alpha+\beta}{2\alpha\beta}\la$. 
From  \eqref{31} and \eqref{41} one can see  
that $\nec$ can be decomposed into $\rab$, while $\necc$ can be decomposed into $\rabb$. 

(2) We discuss the NcHO and the 2pQRM using path measures.
As mentioned above, the 2pQRM is a physical model whose coefficients are expressed as 
$2\times 2$ Pauli matrices, making it suitable for path integrals.
In \cite{HS24}, the Feynman-Kac formula for 
$e^{-t\rabb}$ was constructed. 
Similarly, we construct a Feynman-Kac formula for 
$e^{-t\rab}$ in Theorem~\ref{main2}.
However, since the interaction term of the 1pQRM is first-order with respect to the annihilation  and  creation operator, whereas the interaction term of the 2pQRM is second-order. Therefore,  
constructing the Feynman-Kac formula for $e^{-t\rab}$  is nontrivial.
On the other hand, the NcHO is a mathematical model and appears somewhat artificial.
As a result, it has been challenging to apply the theory of path integrals to 
$e^{-t\nec}$  except in special cases such as when $\alpha=\beta$ \cite{tan08}.
However, in this paper, we elucidate the relationship between the 2pQRM and the NcHO, as shown in \eqref{31}, and derive a path integral for 
$e^{-t\nec}$ based on \eqref{31}.

(3) We discuss spectral zeta functions. 
Let $\sss(\rab)=\{\mu_n(\triangle,g)\}$. Here $\sss(K)$ denotes the spectrum of a self-adjoint operator $K$. Then the spectral zeta function is defined by 
$$\zeta_{2p}(s;\tau)=\sum_{n=0}^\infty\frac{1}{\mu_n(\triangle,g)^s},\quad s>1$$
We study the asymptotic behaviors of the spectral zeta function of 
$\rab$ using the Feynman-Kac formula. 
A similar asymptotic behaviors of the spectral zeta function of 
$\rabb$ have been investigated in \cite{RW23,HS24} using path measures.

We organize this paper as follows. 

In Section \ref{2} 
we review the spectra of $\rab$ and $\nec$, and 
also show that the lowest eigenvalue curves of $\rab$, $\rabb$ and $\nec$ are concave. 
In Section \ref{fib}
we show a fiber decomposition of the NcHO in terms of the 2pQRM. 
In Section \ref{3} we derive Feynman-Kac formulas of the semigroups generated by 
the NcHO and the 2pQRM, and study the asymptotic behaviors of their spectral zeta functions. 
Section \ref{4} is devoted to introducing  $\necc$. 
We establish relationships between $\necc$ and 
 $\rabb$ in Theorem~\ref{main4}. 
We include appendices, Sections \ref{5} and \ref{7}, in this paper.
In Section \ref{5} we review the spectrum of the form $(p+tq)^2+sq^2$ for $t,s\in\RR$, 
which is used in Section \ref{2}. 
In Section \ref{7}  we investigate 
the asymptotic behavior of the spectral zeta function of the one-particle non-commutative harmonic oscillator.

\section{Spectrum of NcHO and 2pQRM}
\label{2}
\subsection{Preliminaries}
Let 
$a$ and $\add$ be the annihilation operator 
and the creation operator in $\LR$, respectively, which are given by 
$
 a=\frac{1}{\sqrt 2} \left(\frac{\rd}{\rd x}+ x\right)$ and
 $ \add=\frac{1}{\sqrt2} \left(- \frac{\rd}{\rd x}+ x\right)$. 
They satisfy the canonical commutation relation 
$[a,\add]=\one$. 
Note that 
\[\add a=-\half\frac{\rd^2}{\rd x^2}+\half x^2-\half\]
is the harmonic oscillator, and $\sss(\add a)={\NN}\cup \{0\}$. 
Let $L_n$ be the eigenspace of $\add a $ for the eigenvalue $n$. 
It is actually given by $L_n={\rm LH}\{\frac{1}{\sqrt n}{\add}^n \gr\}$, where 
$\gr(x)=\pi^{-1/4} e^{-x^2/2}$. Note that $a\gr=0$. 
Then $\dim L_n=1$ for any $n\geq0$ and 
$\LR$ can be represented as a Fock space: 
\[\LR=\bigoplus_{n=0}^\infty L_n.\]
Thus $a:L_n\to L_{n-1}$ for $n\geq 1$ and $\add : L_n\to L_{n+1}$ for $n\geq0$. 
On the other hand $\add a$ leaves $L_n$ invariant. 
Let \[\rU_t=e^{-it \add a}.\] $\{\rU_t\}_{t\in\RR} $ is a strongly continuous one-parameter unitary group and leaves $L_n$ invariant for each $n\geq0$. 
It can be shown that 
$\rU_t$ maps $\rD(a)$ to itself and 
$\rU_ta \rU_t^{-1}=e^{it}a$ on $\rD(a)$, 
and 
$\rU_t$ maps $\rD(\add)$ to itself and 
$\rU_t\add \rU_t^{-1}=e^{-it}\add$ on $\rD(\add)$. 
In particular $e^{-i(\pi/2)\add a}$ is the Fourier transform on 
$\bigoplus_{n=0}^\infty L_n$. 
We set 
\[p=-i\frac{\rd}{\rd x},\quad q=x.\]
Then 
$e^{-i(\pi/2)\add a} (a+\add) e^{i(\pi/2)\add a}=i(a-\add)$, which implies that 
\[e^{-i(\pi/2)\add a} qe^{i(\pi/2)\add a}=-p.\] 
Moreover 
$e^{-i(\pi/4)\add a}$ is very useful in this paper. 
It satisfies that 
\[
e^{-i(\pi/4)\add a} (a^2-{\add}^2) e^{i(\pi/4)\add a}=i(a^2+{\add}^2)\]
which implies that 
\[e^{-i(\pi/4)\add a} (pq+qp) e^{i(\pi/4)\add a}=-p^2+q^2.\]
Next we explain the spin part. 
Let $\s_x$, $\s_y$ and $\s_z$ be the $2\times 2$ Pauli matrices given by 
\begin{align*}
 \s_x= \begin{pmatrix} 0 & 1 \\ 1 & 0 \end{pmatrix},\quad
 \s_y = \begin{pmatrix} 0 & -i \\ i & 0 \end{pmatrix},\quad
 \s_z = \begin{pmatrix} 1 & 0 \\ 0 & -1 \end{pmatrix}.
\end{align*}
The set $\{\s_x,\s_y,\s_z\}$ forms a basis for the Lie-algebra ${\mathfrak s}{\mathfrak u}(2)$. 
The $3$-dimensional rotation group $SO(\RR^3)$ 
has an adjoint representation on $SU(2)$.
Let $n\in \RR^3$ be a unit vector and let $\theta\in [0,2\pi)$. 
Define $\boldsymbol{\sigma}=(\s_x,\s_y,\s_z)$, and for $v=(v_1,v_2,v_3)\in\RR^3$, 
we set  $n\cdot \boldsymbol{\sigma}=n_1\s_x+n_2\s_y+n_3\s_z$. 
It follows that $e^{(i/2)\theta n\cdot \s}(v\cdot \boldsymbol{\sigma}) e^{-(i/2)\theta n\cdot \s}=
Rv\cdot \boldsymbol{\sigma}$, where 
$R\in SO(3)$ denotes the $3\times 3$ matrix representing 
the rotation around $n$ by an angle $\theta$.
As a specific example, let $n=(0,1,0)$ and $\theta=\pi/2$. In this case 
we have 
$\rs_y \s_y \rs_y^{-1}=\s_y$ and 
$\rs_y \s_x \rs_y^{-1}=\s_z$ and 
$\rs_y\s_z \rs_y^{-1}=-\s_x$, 
where 
$\rs_y=e^{i\frac{\pi}{4}\s_y}$. 
In the same way we can construct $\rs_x$ and $\rs_z$. 
In this paper, we frequently use such unitary transformations 
\begin{align}
\label{s1}&\rs_x\s_x\rs_x^{-1}=\s_x,\,\, \rs_x\s_y\rs_x^{-1}= \s_z,\,\, \rs_x\s_z\rs_x^{-1}=-\s_y, \\
\label{s2}&\rs_y\s_x\rs_y^{-1}= \s_z,\,\, \rs_y\s_y\rs_y^{-1}=\s_y, \,\, \rs_y\s_z\rs_y^{-1}=-\s_x, \\
\label{s3}&\rs_z\s_x\rs_z^{-1}=-\s_y,\,\, \rs_z\s_y\rs_z^{-1}= \s_x, \,\, \rs_z\s_z\rs_z^{-1}=\s_z
\end{align}
on spins in the following sections. Combining $\rU_t$ and $\rs_{\cdot}$, we have the unitary equivalences:
\[\s_y(-i)\otimes(a-\add)\cong \s_x\otimes(a+\add),\quad 
\s_y(-i)\otimes(a^2-{\add}^2)\cong \s_x\otimes(a^2+{\add}^2).\]
We shall use these equivalences in this paper. 

\subsection{Two-photon quantum Rabi model}
The state space for the 2pQRM is given by the Hilbert space:
\[\cH=\CC^2\otimes\LR.\] 
We use identifications: 
\[\cH\cong \LR\oplus \LR\cong L^2(\RR;\CC^2)\]
unless it causes confusion. 
The 1pQRM describes 
a two-level atom coupled to a single photon, which 
is introduced by I.I.Rabi \cite{rab36} in the semiclassical region 
and then 
the quantized version is introduced by E.T. Jaynes and F.W. Cumming \cite{JC63}.  
The Hamiltonian is given by 
\[\rabb=\triangle\s_z\otimes \one+\one\otimes 
\lk \add a+\half\rk+ g\s_x\otimes (a+{\add})\]
as a self-adjoint operator in $\cH$, where
$g\in\RR$ stands for a coupling constant, 
and $\triangle>0$ is also a constant. 
In physics $\sss(\tri\s_z)=\{-\tri,\tri\}$ describes the eigenvalues of the two-level atom. 
While the Hamiltonian of the 2pQRM, which is the main subject in this paper, is defined 
by $\rabb$ with the interaction term $a+\add$ replaced by $a^2+{\add}^2$: 
\begin{align}\label{rabi}
\rab =\triangle\s_z\otimes \one+\one\otimes 
\left( \add a+\half\right)+ 
g\s_x\otimes (a^2+{\add}^2). 
\end{align}
In physics 
the 2pQRM is defined by 
$\rab $ with 
$\add a+\half$ replaced by 
$\add a$, but in this paper 
we add an extra term $\half$ and adopt \eqref{rabi} for the latter convenience. 
Notice that 
\[a^2+{\add}^2=\frac{\rd^2}{\rd x^2}+ x^2=- p^2+ q^2.\]
For $\one\otimes\add a$, 
we use identifications:
\[\one\otimes\add a=\add a \oplus \add a=(\half p^2+\half q^2+\half)\oplus (\half p^2+\half q^2+\half).\]
Let $\cD=\rD(\one\otimes\add a)$. 
We also identify $\cD$ with several subspaces below:
\[\cD=\CC^2\otimes \rD(\add a)=\rD(\add a)\oplus \rD(\add a)=\rD(\add a\oplus \add a)=
\rD(p^2+q^2)\oplus \rD(p^2+q^2).\]
We use these identifications unless any confusion arises. 
We emphasize that 
\[\rD(\add a)=\rD\left( \frac{\rd ^2}{\rd x^2}\right) \cap \rD(x^2).\] 

The proposition below shall be often times used in this paper. 
\begin{proposition}\label{p}
Let $t,s\in\RR$ and $T_{t,s}=(p+tq)^2+sq^2$. 
\begin{description}
\item[$(s>0)$] $T_{t,s}$ is self-adjoint on $\rD(p^2+q^2)$, $T_{t,s}\cong \sqrt{s}(p^2+q^2)$ and 
$\sss(T_{t,s})=\{\sqrt s(2n+1)\}$. 
\item[$(s=0)$] $T_{t,s}$ is self-adjoint on $\cU_t\rD(p^2)$, $T_{t,s}\cong p^2$ and $\sss(T_{t,s})=[0,\infty)$, where $\cU_t=e^{itq^2/2}$. 
\item[$(s<0)$] $T_{t,s}$ is essentially self-adjoint on $\rD(p^2+q^2)$, $T_{t,s}\cong \sqrt{-s}(p^2-q^2)$ 
and $\sss(T_{t,s})=\RR$. 
\end{description}
\end{proposition}
\proof
See Theorems \ref{s=0}, \ref{s>0} and \ref{s<0} in Section \ref{5}. 
\qed
Let $g=0$. Then we see that 
\begin{align*}
&H_{\tri,0}^{\rm 2p}=\M{\half(p^2+q^2)+\tri&0\\0&\half(p^2+q^2)-\tri},\\
&\sss(H_{\tri,0}^{\rm 2p})=\{n+\half\pm\tri\}.
\end{align*}
On the other hand, by the definition of $\rab$ in \eqref{rabi} and the fact $\sss(a^2+{\add}^2)=\RR$, 
the spectrum of $\rab$ for $g\neq 0$ drastically varies according to the value of $g$. 
Now let us consider 
$h_\eps=\half p^2+\eps\half q^2$. 
The spectrum of $h_\eps$ drastically varies in accordance with parameter $\eps$:
\begin{align*}
\sss(h_\eps)=\left\{
\begin{array}{ll}
\{\sqrt\eps(n+1)/2\}_{n=0}^\infty& \eps>0,\\ 
{[0,\infty)}& \eps=0,\\ 
\RR& \eps<0
\end{array}
\right.
\end{align*}
by Proposition \ref{p}. $\sss(\rab)$ also has a similar property. 
Now we consider the case of $\tri=0$. 

\begin{lemma}[spectrum of $H_{0,g}^{\rm 2p}$]\label{re}
$H_{0,g}^{\rm 2p}$ is self-adjoint on 
$\rD(p^2+q^2)\oplus \rD(p^2+q^2)$ for $|g|<\half$,
self-adjoint on $\rD(q^2)\oplus \rD(p^2)$ for $|g|=\half$,
and essentially self-adjoint on 
$\rD(p^2+q^2)\oplus \rD(p^2+q^2)$ for $|g|>\half$. 
Moreover we have
\begin{align*}
&H_{0,g}^{\rm 2p}\cong \lkk\begin{array}{ll}
\half \sqrt{1-4g^2}\M{p^2+q^2&0\\0&p^2+q^2}&|g|<\half,\\
\M{q^2&0\\0&p^2}& |g|=\half,\\
\half \sqrt{4g^2-1}\M{-p^2+q^2&0\\0&p^2-q^2}&|g|>\half.
\end{array}\right.
\end{align*}
In particular 
\begin{align*}
&\sss(H_{0,g}^{\rm 2p})=
\lkk
\begin{array}{ll} 
\sqrt{1-4g^2}\{ n+\half\}&|g|<\half,\\ 
{[0,\infty)}& |g|=\half,\\
\RR&|g|>\half.
\end{array}
\right.
\end{align*}
\end{lemma}
\proof
We have 
\begin{align*}
\rs_y H_{0,g}^{\rm 2p}\rs_y^{-1}
 =\M{
(\half-g)p^2+(\half+g)q^2&0\\
0& (\half+g)p^2+(\half-g)q^2}.
\end{align*}
Then the lemma follows from Proposition \ref{p}. 
\qed
Now we consider the case of $\tri\neq0$. 

\begin{lemma}[self-adjointness of $\rab$]\label{ess}\ 
\begin{description}
\item[$(|g|<\frac{1}{2})$] 
$\rab $ is self-adjoint on $\rD(p^2+q^2)\oplus \rD(p^2+q^2)$ 
and bounded from below. $\sss(\rab)$ is purely discrete and 
\begin{align}\label{gr}
\inf\sss(\rab)\geq \half \sqrt{1-4g^2}-\triangle.
\end{align}
\item[$(|g|=\frac{1}{2})$]
$\rab $ is self-adjoint on $\rD(q^2)\oplus \rD(p^2)$ 
and 
\begin{align}\label{gr2}
\inf\sss(\rab)\geq-\triangle.
\end{align}
\item[$(|g|<\frac{1}{2})$]
$\rab $ is essentially self-adjoint on $\rD(p^2+q^2)\oplus \rD(p^2+q^2)$ 
and \[\inf \sss(\rab)=-\infty.\] 
\end{description}
\end{lemma}
\proof
Suppose that $|g|<\half$. 
By the unitary transformation $\rs_y$ in \eqref{s2}, 
\begin{align}\label{g}
\rs_y \rab \rs_y ^{-1}=
\M{
(\half-g)p^2+(\half+g)q^2&-\triangle\\
-\triangle& (\half+g)p^2+(\half-g)q^2}.
\end{align}
Since $\rs_y \rab \rs_y ^{-1}=
\rs_y H_{0,g}^{\rm 2p}\rs_y^{-1} -\tri\s_x$ and 
$-\tri\s_x$ is bounded and self-adjoint, 
$\rs_y \rab \rs_y ^{-1}$ is self-adjoint on 
$\rD(p^2+q^2)\oplus \rD(p^2+q^2)$ 
and bounded from below by 
Lemma \ref{ess} and Kato-Relich theorem \cite{kat51}. 
$\s_x(\rs_y H_{0,g}^{\rm 2p}\rs_y^{-1}+a)^{-1}$ is compact for any $a>0$. 
Then $\sss(\rab)$ is purely discrete. 
\eqref{gr} follows from
$\half \sqrt{1-4g^2}\leq H_{0,g}^{\rm 2p}$
and 
\begin{align}\label{ii}
-\triangle+H_{0,g}^{\rm 2p}
\leq 
\rab
\leq \triangle +H_{0,g}^{\rm 2p}.
\end{align}
Proofs for the cases $|g|=\half$ and $|g|>\half$ are similarly established by using \eqref{ii} and Proposition~ \ref{p}. 
\qed

We consider the parity symmetry. 
The parity symmetry 
of $\rabb$ is very useful for studying the spectrum of $\rabb$. 
For $\rab$ there is a similar symmetry. 
Let 
$P_1=\s_z\otimes e^{i\pi {\add a}}$ and 
$P_2=\s_z\otimes e^{i(\pi/2) {\add a}}$.
It can be verified that 
$[P_1,\rabb]=0$ and $[P_2,\rab]=0$. 
The former symmetry is known as $\ZZ_2$-symmetry or parity symmetry, 
while the latter is referred to as $\ZZ_4$-symmetry. 
Since the spectrum of $P_2$ is $\{\pm 1,\pm i\}$, 
the state space $\cH$ can be decomposed into 
four subspaces:
\[\cH=
\cH_{+1}\oplus \cH_{-1}\oplus\cH_{i}\oplus\cH_{-i},\]
where 
$\cH_k$ is the eigenspace of $P_2$ corresponding to eigenvalue $k\in \{\pm1,\pm i\}$. 
Let $\CC^2\otimes L_n=L_{n+}\oplus L_{n-}$, where 
$L_{n+}=\{\binom{f}{0}\mid f\in L_n\}$ and 
$L_{n-}=\{\binom{0}{f}\mid f\in L_n\}$.
Then each $\cH_k$ is defined by 
\begin{align*}
&\cH_{+1}=\bigoplus_{m=0}^\infty L_{4m+}\oplus L_{4m+2-},\quad \cH_{-1}=\bigoplus_{m=0}^\infty L_{4m+2+}\oplus L_{4m-},\\
&\cH_{+i}=\bigoplus_{m=0}^\infty L_{4m+1+}\oplus L_{4m+3-},\quad \cH_{-i}=\bigoplus_{m=0}^\infty L_{4m+3+}\oplus L_{4m+1-}.
\end{align*}
$\rab$ can be reduced by $\cH_k$. 
In the case of $g=0$ it can be observed that the ground state $\binom{0}{\gr}$ belongs to  $\cH_{-1}$. 
We shall show  in Corollary \ref{parity}, 
using path measures,  that the ground state $\Phi$ of $\rab$ satisfies 
$\Phi\in \cH_{-1}$ for any $g\in\RR$.

\subsection{Non-commutative harmonic oscillator}
In this section we are concerned with the NcHO. 
Let $\alpha,\beta\in\RR$. The NcHO is a self-adjoint operator 
in 
$\cH$, 
which is defined by 
\[\nec =\AB\otimes \left( 
\add a +\half\right) +\half\J \otimes (a^2-{\add}^2).\]
We notice that the interaction term in $\nec$ is 
\[\half(a^2-{\add}^2)
=\half\left( x\frac{\rd}{\rd x}+\frac{\rd}{\rd x} x\right)=
x\frac{\rd}{\rd x}+\half=i\half (pq+qp).
\]
Note that $\sss((-i)(a^2-{\add}^2))=\RR$. 

\begin{lemma}[essential self-adjointness of $\nec$]\label{qesa}
Let $\alpha,\beta\in\RR$.
Then $\nec$ is essentially self-adjoint on any core of $\one\otimes \add a$. 
\end{lemma}
\proof
We have 
\[\nec =\AB\otimes \half(p^2+q^2)+ \s_y\otimes \half(pq+qp).\]
Note that 
$\nec$ is symmetric on $\cD$.
Let $N=\one\otimes(\add a+1)$. 
Thus directly we see that 
$[N,\nec]=2i\s_y\otimes (q^2-p^2)$.
Then we have 
\begin{align*}
&|(f, \nec g)|\leq C\|N^\frac{1}{2} f\|\|N^\frac{1}{2} g\|,\\
&|(Nf, \nec g)-
(\nec f, Ng)|=2|(f,i\s_y \otimes (q^2-p^2)g)|
\leq C\|N^\frac{1}{2} f\|\|N^\frac{1}{2} g\|.
\end{align*}
By the Nelson commutator theorem \cite[Theorem X.36']{RS2}, 
$\nec $ is essentially self-adjoint on any core of $\one\otimes \add a$.
\qed
In the special case where $\alpha,\beta>0$
 we can 
 provide further details
on the spectrum of $\nec$. 
Let 
\[\AAA =\M{\sqrt\alpha&0\\0&\sqrt\beta}\otimes\one.\]
Then 
$\AAA \cD =\cD $ 
and 
\begin{align}\label{A}
\nec=\AAA \nek \AAA \end{align}
on $\cD $, where 
\[\nek=\one \otimes \half(p^2+q^2)+
\frac{1}{2\sqrt{\alpha\beta}}\s_y \otimes (pq+qp).\]
The following lemmas (Lemmas \ref{K} and \ref{qsa}) may be well-known, but we include their proofs for the sake of self-consistency, as the proofs are short and fundamental

\begin{lemma}[spectrum of $\nek$]\label{K}
Operator $\nek$ is self-adjoint on 
$\rD(p^2+q^2)\oplus\rD(p^2+q^2)$ for 
$\alpha\beta>~1$,
self-adjoint on $\rD((p+q)^2)\oplus \rD((p-q)^2)$ for $\alpha\beta=1$, 
and 
essentially self-adjoint on $\rD(p^2+q^2)\oplus\rD(p^2+q^2)$ for 
$0<\alpha\beta<1$. 
Moreover we have 
\[\nek\cong\lkk\begin{array}{ll}
\half \sqrt{1-\frac{1}{\alpha\beta}}
\M{ p^2+ q^2
&0\\
0& p^2+ q^2}& \alpha\beta>1,\\
\half \M{ (p+q)^2
&0\\0& (p-q)^2}
&\alpha\beta=1,\\
\half\sqrt{\frac{1}{\alpha\beta}-1}
\M{ p^2- q^2
&0\\0& p^2- q^2
}&0<\alpha\beta<1.
\end{array}\right.\]
In particular 
\[\sss(\nek)=\lkk\begin{array}{ll}
\sqrt{1-\frac{1}{\alpha\beta}}\left\{n+\half\right\}_{n=0}^\infty& \alpha\beta>1,\\
\sss(\nek)=[0,\infty)& \alpha\beta=1,\\
 \sss(\nek)=\RR&0<\alpha\beta<1.\end{array}
 \right.\] 
\end{lemma}
\proof
Suppose that $\alpha\beta>1$. 
We see that 
\[\nek=\half\left(
\one\otimes p+
\frac{1}{\sqrt{\alpha\beta}}\s_y\otimes q\right)^2+\half\left(1-\frac{1}{\alpha\beta}\right) \one\otimes q^2.\]
By $\rs_x$ in \eqref{s1} we have
\begin{align*}
\rs_ x\nek\rs_x^{-1}&=
\half\left(
\one\otimes p+\frac{1}{\sqrt{\alpha\beta}}\s_z \otimes q\right)^2+\half\left(1-\frac{1}{\alpha\beta}\right) \one\otimes q^2\\
&=\half 
\M{
\left(p+\frac{1}{\sqrt{\alpha\beta}}q\right)^2+\left(1-\frac{1}{\alpha\beta}\right) q^2&0\\
0&
\left(p-\frac{1}{\sqrt{\alpha\beta}}q\right)^2+\left(1-\frac{1}{\alpha\beta}\right) q^2
}.\end{align*}
$\left(p\pm\frac{1}{\sqrt{\alpha\beta}}q\right)^2+
\left(1-\frac{1}{\alpha\beta}\right) q^2$ 
is self-adjoint on $\rD(p^2+q^2)$ by Proposition \ref{p}. Then 
$\nek$ is self-adjoint on $\rD(p^2+q^2)\oplus \rD(p^2+q^2)$. 
Moreover 
\begin{align*}
\rs_ x\nek\rs_x^{-1}
&
\cong\half
\M{ p^2+\left(1-\frac{1}{\alpha\beta}\right) q^2
&0\\0& p^2+\left(1-\frac{1}{\alpha\beta}\right) q^2}\\
&\cong 
\half \sqrt{1-\frac{1}{\alpha\beta}}
\M{ p^2+ q^2
&0\\
0& p^2+ q^2}
\end{align*}
by Proposition \ref{p}. 
Then 
$\sss(\nek)=\left\{\sqrt{1-\frac{1}{\alpha\beta}}\lk n+\half\rk\right\}_{n=0}^\infty$. 
Suppose that $\alpha\beta=1$. 
Then 
\begin{align*}
\rs_x \nek\rs_x^{-1}&=
\half\M{ (p+q)^2
&0\\0& (p-q)^2
}
\end{align*}
and $\nek$ is self-adjoint on $\rD((p+q)^2\oplus (p-q)^2)$ with 
$\sss(\nek)=[0,\infty)$ by Proposition \ref{p}. 
Finally 
suppose that $0<\alpha\beta<~1$. 
By Proposition \ref{p} again we see that $\nek$ is essentially self-adjoint on $\rD(p^2+q^2)\oplus \rD(p^2+q^2)$. 
Moreover we have 
\begin{align*}
\nek&\cong
\half \M{ p^2+\left(1-\frac{1}{\alpha\beta}\right) q^2
&0\\0& p^2+\left(1-\frac{1}{\alpha\beta}\right) q^2}\\
&\cong
\half\sqrt{\frac{1}{\alpha\beta}-1}
\M{ p^2- q^2
&0\\0& p^2- q^2
}.
\end{align*}
Then 
$\sss(\nek)=\RR$. 
\qed

\begin{lemma}[self-adjointness of $\nec$]\label{qsa}\ 
\begin{description}
\item[($\alpha\beta>1$)]
$\nec$ is self-adjoint on 
$\rD(p^2+q^2)\oplus\rD(p^2+q^2)$ and $\sss(\nek)$ is purely discrete and 
\begin{align}
\label{ei}
\left( n+\half\right)\min\{\alpha,\beta\}
\sqrt{1-\frac{1}{\alpha\beta}}
\leq\la_{2n}\leq \la_{2n+1}\leq
\left( n+\half\right)\max\{\alpha,\beta\}
\sqrt{1-\frac{1}{\alpha\beta}}
\end{align}
for $n\geq0$. 
In particular 
\begin{align*}
\inf\sss(\nec)\geq 
\frac{\min\{\alpha,\beta\}}{2}
\sqrt{1-\frac{1}{\alpha\beta}}. 
\end{align*}
\item[($\alpha\beta=1$)]
$\nec$ is self-adjoint on $\rD((p+q)^2\oplus (p-q)^2)$ and 
 \[\inf\sss(\nec)=0.\] 
\item[($0<\alpha\beta<1$)]
$\nec$ is essentially self-adjoint on $\rD(p^2+q^2)\oplus\rD(p^2+q^2)$
and 
 \[\inf\sss(\nec)=-\infty.\] 
\end{description}
\end{lemma}
\proof
Suppose that $\alpha\beta>1$. 
It can be seen that $\AAA \nek \AAA$ is symmetric. 
Let $g\in \rD((\AAA \nek \AAA )^\ast)$ and $f\in 
\cD $. Then there exists $h$ such that 
$(\AAA \nek \AAA f,g)=(f,h)$. 
On the other hand 
$(\nek \AAA f,\AAA g)=(f,h)$. 
Since 
$\AAA \cD =\cD $, 
$(\nek \AAA f,\AAA g)=(\AAA f,\AAA ^{-1}h)$ implies that $\AAA g\in \rD(\nek )$ and 
$\nek \AAA g=\AAA ^{-1}h$.
Hence $\AAA \nek \AAA g=h$ and $g\in \rD(\AAA \nek \AAA )$. 
Thus $\rD(\AAA \nek \AAA )=\rD((\AAA \nek \AAA )^\ast)$. Then $\AAA \nek \AAA$ is self-adjoint
on $\cD $. 
Let $\sss(\nek )=\{e_n\}$. 
By the minmax principle 
we have the $n$th eigenvalue of $\nec$ is given by 
\begin{align*}
\la_n&=
\sup_{f_0,\ldots,f_n\in \rD(\nec)}\inf_{f\in \{{\rm LH}[f_0,\ldots,f_n]\}^\perp}
\frac{(f,\nec f)}{\|f\|^2} \\
&=
\sup_{f_0,\ldots,f_n\in \rD(\AAA \nek \AAA )}
\inf_{f\in \{{\rm LH}[f_0,\ldots,f_n]\}^\perp}
\frac{\|\AAA f\|^2}{\|f\|^2}\frac{(\AAA f,\nek \AAA f)}{\|\AAA f\|^2} 
\geq 
\min\{\alpha,\beta\} e_n.
\end{align*}
Similarly we also have 
$\la_n\leq
\max\{\alpha,\beta\} e_n$. 
Since $e_n\to\infty$ as $n\to\infty$, 
$\la_n\to\infty$ as $n\to\infty$. Thus 
$\sss(\nec)$ is purely discrete, and \eqref{ei} follows. 
Finally we estimate the lowest eigenvalue $\la_0$ of $\nec$. 
Since 
\[\la_0
=\inf_{f\in \rD(\AAA \nek \AAA )}\frac{(\AAA f, \nek \AAA f)}{\|\AAA f\|^2}\frac{\|\AAA f\|^2}{\|f\|^2}\geq
\min\{\alpha,\beta\}\inf\sss(\nek ),\]
hence \eqref{ei} follows. 
Next suppose that $\alpha\beta=1$. 
The self-adjointness can be proven in a similar way to the case of $\alpha\beta>1$. 
Note that $\AAA$ is invertible. There exists $\{f_n\}$ such that $(f_n, \nek f_n)\to 0$ as $n\to\infty$. 
Let $\AAA^{-1}f_n=g_n$. Then 
 $(g_n, \nec g_n)\to 0$ as $n\to\infty$ 
 and 
 $\inf\sss(\nec)=0$ follows. 
Finally suppose that $0<\alpha\beta<1$. We also see that 
there exists $f_n$ such that $\|f_n\|=1$ and 
$(f_n, \nec f_n)=(\AAA f_n, \nek \AAA f_n)\to -\infty$ as $n\to\infty$ by Lemma~\ref{K}. 
Then $\inf\sss(\nec)=-\infty$ follows. 
\qed

\begin{remark}
\bi
\item[(1)] An upper bound of $\la_0$ is also derived in \cite[Theorem 8.2.1]{par10}:
\[\la_0\leq \frac{\sqrt{\alpha\beta}\sqrt{\alpha\beta-1}}{\alpha+\beta+|\alpha-\beta|{(\alpha\beta-1)^{1/4}}\cos(\half\arctan\frac{1}{\sqrt{\alpha\beta-1}})}.\]
\item[(2)] \eqref{ei} is also proven in \cite{IW07}.
\item[(3)] Let $\alpha\beta=1$. 
Then $\nec =\AAA \nek \AAA$ and $\sss(\nek )=[0,\infty)$, 
but it is not trivial to show $\sss(\nec)=[0,\infty)$, since $\AAA$ is not unitary. 
It is however shown in \cite{PV13} that 
$\sss(\nec)=[0,\infty)$ in the case of $\alpha\beta=1$. 
\item[(4)] In \cite{PW02a,PW02b} it is shown that 
for 
$\alpha=\beta$ and $\alpha^2\geq1$, 
\begin{align}\label{qe}
Q_{\alpha\alpha}^{\rm NH}\cong \lkk\begin{array}{ll}
\half \sqrt{\alpha^2-1}\M{ p^2+ q^2 
& 0\\ 0& p^2+ q^2}&\alpha^2>1\\
\M{p^2&0\\0&q^2}&\alpha^2=1.
\end{array}\right.
\end{align}
\item[(5)]
Suppose that $\alpha,\beta>0$ and $\alpha\beta>1$. 
Then it is also shown in \cite{PW02a,PW02b}
that $\nec$ and $\cen$ are unitarily equivalent. 
It is actually given by 
\begin{align}\label{s}
e^{-i(\pi/2)\add a}\s_x  \nec  \s_x ^\ast e^{i(\pi/2)\add a}
=\cen.
\end{align}
\ei
\end{remark}

\subsection{Concavity of the lowest eigenvalue curves}
\label{6}
In this section we show that the lowest eigenvalue curves of $\rab$, $\rabb$ and $\nec$ are 
concave. More precisely, the lowest eigenvalue $E$ of these models is a function of 
the coupling constant $g$, and we shall show that 
 the function  \[g\mapsto E(g)\] is concave. 
 For $\rabb$ this can be shown using Feynman-Kac formula (see Remark \ref{con}), 
 but 
 for $\rab$ and $\nec$ it is hard to show this in a similar manner to $\rabb$. 
 We begin with demonstrating  perturbative computations of the lowest eigenvalues of $\rab$, 
 $\rabb$ and $\nec$, 
since these models are analytic family of type (A) \cite[p.16]{RS4} under certain conditions. 
Consequently the lowest eigenvalues and their corresponding eigenvectors are analytic in 
the coupling constant. 
A similar result for 
perturbative computations of the lowest eigenvalues of $\nec$ is given in \cite[Section 3.2]{par04}. 

An abstract procedure for the computation is as follows. 
Let $H(0)$ be a self-adjoint operator and $V$ is symmetric. 
Define $H(g)=H(0)+gV$ for $g\in\RR$ such that 
$H(g)$ is an analytic family of type (A). 
Set $\inf\sss(H(g))=E(g)$ and $H(g)\Phi(g)=E(g)\Phi(g)$ with $\|\Phi(g)\|=1$ for any $g\in\RR$. 
Suppose that $E(g)$ is discrete and simple for any $g\in\RR$. 
Then Kato-Rellich theorem yields that $E(g)$ is analytic in $g$ and $\Phi(g)$ is also analytic in $g$ \cite[Theorem XII.8]{RS4}. 
Suppose that $E(g)=E(-g)$ and 
$H(0)\Phi_0=E(0)\Phi_0$. 
Then we have 
\begin{align}
\label{e}
E^{(2)}(0)=-2(V\Phi_0, (H(0)-E(0))^{-1} V\Phi_0),
\end{align}
where  $E^{(n)}$ and $\Phi^{(n)}$ denote the $n$th derivative of $E$ and $\Phi$, respectively. 
 This formula can be derived by considering 
the eigenvalue equation 
$H(g)\Phi(g)=E(g)\Phi(g)$. 
Note that $\Phi(0)=\Phi_0$. 
We have 
\begin{align*}
&V\Phi(g)+(H(0)+gV)\Phi^{(1)}(g)=E^{(1)}(g)\Phi(g)+E(g)\Phi^{(1)}(g),\\
&2V\Phi^{(1)}(g)+(H(0)+gV)\Phi^{(2)}(g)=E^{(2)}(g)\Phi(g)+2E^{(1)}(g)\Phi^{(1)}(g)+E(g)\Phi^{(2)}(g).
\end{align*} 
By $E^{(1)}(0)=0$, it follows that $E^{(2)}(0)=(2V\Phi^{(1)}(0), \Phi_0)$. 
\eqref{e} follows from $\Phi^{(1)}(0)=-(H(0)-E(0))^{-1}V\Phi_0$. 
This procedure can be extended straightforwardly as follows. 
Set $\Phi(0)=\Phi$, $\Phi^{(n)}(0)=\Phi^{(n)}$, $E^{(n)}(0)=E^{(n)}$
and $H(0)^{-1}=\frac{1}{K}$, and assume that $E(0)=E^{(2m+1)}(0)=0$ for $m\geq0$. 
Then 
\begin{align*}
&nV\Phi^{(n-1)}+H(0)\Phi^{(n)}=\sum_{k=1}^{[n/2]}\binom{n}{2k}E^{(2k)}\Phi^{(n-2k)},\\
&\Phi^{(n)}=\frac{1}{K}\lk -nV\Phi^{(n-1)}+\sum_{k=1}^{[n/2]}\binom{n}{2k}E^{(2k)}\Phi^{(n-2k)}
\rk.
\end{align*}
The first several terms are given by 
\begin{align*}
&2(V\Phi,\Phi^{(1)})=E^{(2)},\\
&3(V\Phi,\Phi^{(2)})=\binom{3}{2}E^{(2)}(\Phi,\Phi^{(1)}),\\
&4(V\Phi,\Phi^{(3)})=\binom{4}{2}E^{(2)}(\Phi,\Phi^{(2)})+E^{(4)}
\end{align*}
and 
\begin{align*}
&\Phi^{(1)}=-\frac{1}{K}V\Phi,\\
&\Phi^{(2)}=
\frac{1}{K}(-2V\Phi^{(1)}+E^{(2)}\Phi)=
\frac{1}{K}Y\Phi,
\\
&\Phi^{(3)}=
\frac{1}{K}(-3V\Phi^{(2)}+3E^{(2)}\Phi^{(1)})=
-3\frac{1}{K}V\frac{1}{K}Y\Phi-3E^{(2)}\frac{1}{K}\frac{1}{K}
V\Phi, 
\end{align*}
where 
$Y=2V\frac{1}{K}V+E^{(2)}$. 
Together with \eqref{e} we see that 
\begin{align}
\label{eee}
&E^{(2)}=-2(V\Phi, \frac{1}{k}V\Phi),\\
&E^{(4)}=4(V\Phi, \Phi^{(3)})-6E^{(2)}(\Phi, \Phi^{(2)})
=-6(Y\Phi, \frac{1}{K}Y\Phi)
-12E^{(2)}\| \frac{1}{K}V\Phi\|^2\label{ee}. 
\end{align}
\begin{remark}
We introduce notation $T_\ren=T-\lr{T}$, where $\lr T=(\Phi, T\Phi)$. 
Then \[\frac{1}{2!}E^{(2)}=-\lr{V\frac{1}{K}V}.\]
Notice that $V\frac{1}{K}V\Phi\not\in \rD(\frac{1}{K})$ but 
$(V\frac{1}{K}V)_\ren\Phi\in \rD(\frac{1}{K})$. Thus 
\[\frac{1}{4!}E^{(4)}=-\lr{(V\frac{1}{K}V)_\ren\frac{1}{K}(V\frac{1}{K}V)_\ren}+
\lr{V\frac{1}{K}V}\lr{V\frac{1}{K^2}V}\] is well defined. 
\end{remark}

\subsubsection{2pQRM}
We apply \eqref{eee} and \eqref{ee} to compute the Taylor expansion of the lowest eigenvalue of $\rab$. 
We see that $\rab$ is an analytic family of type (A) for $|g|<\half$ by Lemma \ref{ess}. 
Let $e_\tri^{\rm 2p}(g)$ be the lowest eigenvalue of $\rab$. 
We shall show that $e_\tri^{\rm 2p}(g)$ is simple in Corollary \ref{nocrossing}. 
At $g=0$ we observe that $e^{\rm 2p}_\tri(0)=\half -\tri$ and 
the expansion is given by 
\[e_0^{\rm 2p}(g)=\half\sqrt{1-4g^2}=\half-g^2-g^4+\cO(g^5).\] 
Using \eqref{eee} we can compute 
${e_\tri^{\rm 2p}}^{(2)}(0)$. Let $\Phi=\binom{0}{\gr}$. Then 
$H_{\tri,0}^{\rm 2p}\Phi=(\half -\tri) \Phi$. 
The result is 
\begin{align}
{e^{\rm 2p}_\tri}^{(2)}(0)
&=-2(\s_x\otimes (a^2+{\add}^2)\Phi, (\tri\s_z\otimes\one+\one\otimes\add a+\tri)^{-1}
\s_x\otimes (a^2+{\add}^2)\Phi)_{\cH}\nonumber\\
&=-2\lk \binom{{\add}^2\gr}{0}, \M{(\add a+2\tri)^{-1}& 0\\ 0&(\add a)^{-1}}
\binom{{\add}^2\gr}{0}\rk_{\cH} \nonumber\\
&\label{d1}=-2({\add}^2\gr, (\add a+2\tri)^{-1}{\add}^2\gr)_{\LR}
=-\frac{2}{2+2\tri} \|{\add}^2\gr\|_{\LR}^2=-\frac{2}{1+\tri}.
\end{align}
Similar result can be found in e.g., \cite[(34)]{tra12}.
By setting 
$V=\s_x\otimes (a^2+{\add}^2)$, 
$E^{(2)}=\frac{-2}{1+\tri}$, 
$\Phi=\binom{0}{\gr}$
and 
$\frac{1}{k}=
\MM{(\add a+2\tri)^{-1}& 0\\ 0&(\add a)^{-1}}
$ in \eqref{ee}, 
we can compute ${e_\tri^{\rm 2p}}^{(4)}(0)$. 
Note that 
\begin{align*}
&2\frac{1}{K}V\Phi=\frac{1}{1+\tri}\binom{{\add}^2\gr}{0},\\
&2V\frac{1}{K}V\Phi=\frac{1}{1+\tri}\binom{0}{({\add}^4+2)\gr},\\
&Y\Phi=2V\frac{1}{K}V\Phi+E^{(2)}\Phi=\frac{1}{1+\tri}\binom{0}{{\add}^4\gr},\\
&\frac{1}{K}Y\Phi=\frac{1}{4}\frac{1}{1+\tri}\binom{0}{{\add}^4\gr}.
\end{align*}
Since $\|{\add}^{2n}\Phi\|^2=n!$, we have 
\begin{align*}
&-12E^{(2)}\|\frac{1}{K}V\Phi\|^2
=12\lk\frac{1}{1+\tri}\rk^3,\\
&-6(Y\Phi, \frac{1}{K}Y\Phi)=
-36\lk\frac{1}{1+\tri}\rk^2. \end{align*}
Substituting these terms into \eqref{ee} we can see that 
\begin{align}
\label{d2}
{e^{\rm 2p}_\tri}^{(4)}(0)=-36\lk\frac{1}{1+\tri}\rk^2+12\lk\frac{1}{1+\tri}\rk^3.
\end{align}
\begin{lemma}
We have 
\begin{align}\label{eff}
e^{\rm 2p}_\tri(g)
=
\half -\tri-\frac{1}{1+\tri}g^2-\half \lk\frac{1}{1+\tri}\rk^2\lk \frac{2+3\tri}{1+\tri}\rk g^4+\cO(g^5).
\end{align} 
\end{lemma}
\proof
Since 
$e^{\rm 2p}_\tri(g)
=
e^{\rm 2p}_\tri(0)+\frac{1}{2!}{e^{\rm 2p}_\tri}^{(2)}(0)g^2+
\frac{1}{4!}{e^{\rm 2p}_\tri}^{(4)}(0)g^4+\cO(g^5)$, 
the lemma follows from \eqref{d1} and \eqref{d2}. 
\qed

\begin{theorem}[concavity of eigenvalue curves of $\rab$]
Both of the lowest eigenvalue curves $f_1: g\to e^{\rm 2p}_\tri$ and 
$f_2:g^2\to e^{\rm 2p}_\tri$ are concave for sufficiently small $|g|$.
\end{theorem}
\proof
The coefficient of the $g^2$ term of the expansion of 
$e^{\rm 2p}_\tri$ with respect to $g$ is 
$-\frac{1}{1+\tri}$
 by \eqref{eff}. Then $f_1$ is concave. 
 Similarly 
 the coefficient of the $g^4$ term of the expansion of 
$e^{\rm 2p}_\tri$ with respect to $g^2$ is 
$-\half \lk\frac{1}{1+\tri}\rk^2\lk \frac{2+3\tri}{1+\tri}\rk<~0$. 
Then $f_2$ is also 
 concave. 
\qed

\subsubsection{1pQRM}
We see that $\rabb$ is an analytic family of type (A) for any $g\in\RR$. 
Let $e_\tri^{\rm 1p}(g)$ be the lowest eigenvalue of $\rabb$. 
It is known that $e_\tri^{\rm 1p}(g)$ is simple by \cite{HH14}. 
Let $e^{\rm 1p}_\tri(g)$ be the lowest eigenvalue of $\rabb$. 
Thus $e^{\rm 1p}_0(g)=\half-g^2$. 
In the same way as \eqref{eff} we can also compute ${e^{\rm 1p}_\tri}^{(2)}(0)$. 
The result is 
\begin{align*}
&{e^{\rm 1p}_\tri}^{(2)}(0)
=-2(\s_x\otimes (a+{\add})\Phi, (\tri\s_z\otimes\one+\one\otimes\add a+\tri)^{-1}
\s_x\otimes (a+{\add})\Phi)=-\frac{2}{1+2\tri},\\
&{e^{\rm 1p}_\tri}^{(4)}(0)
=-6(Y\Phi, \frac{1}{K}Y\Phi)
-12E^{(2)}\| \frac{1}{K}V\Phi\|^2
=\frac{-24}{(1+2\tri)^2}+\frac{24}{(1+2\tri)^3}=\frac{-48\tri}{(1+2\tri)^3}.
\end{align*}
Then 
\[e_\tri^{1{\rm p}}(g)=\half -\tri-\frac{1}{1+2\tri}g^2-\frac{2\tri}{(1+2\tri)^3}g^4+\cO(g^5).\] 
By this 
we see that both of the lowest eigenvalue curves $f_1: g\to e^{\rm 1p}_\tri$ and 
$f_2:g^2\to e^{\rm 1p}_\tri$ are concave for sufficiently small $|g|$.
\begin{remark}\label{con}
In \cite[Corollary 4.6]{HS24} it is shown that $f_1$ is concave for all $g\in\RR$ by path measures. 
\end{remark}

\subsubsection{NcHO}
For $\rab$ we could  compute the lowest eigenvalues. 
We can also obtain similar results for  $\nec$. 
We have 
\[\nec\cong \frac{\alpha+\beta}{2}\one\otimes\lk \add a+\half\rk+\s_z\otimes \half(a^2+{\add}^2)-
\frac{\alpha-\beta}{2}\s_y\otimes \lk \add a+\half\rk.\]
We fix $A=\frac{\alpha+\beta}{2}$ and $g=\frac{\alpha-\beta}{2}$. 
Hence 
\begin{align*}
\nec\cong\half\M{A_-p^2+A_+q^2&0\\0&A_+p^2+A_-q^2}-\half g\M{0&-i (p^2+q^2)\\ i(p^2+q^2)&0}
\end{align*}
with $A_+=A+1$ and $A_-=A-1$. 
We regard $A$ as a constant and $g$ a coupling constant. 
Note that $\nec$ is an analytic family of type (A) for sufficiently small $|g|$ by Lemma \ref{qsa}. 
Let $\la_0(g)$ be the lowest eigenvalue of $\nec$. 
It is known that $\la_0(g)$ is simple by \cite{HS13, HS14}. 
By \eqref{s} we see that $\la_0(g)=\la_0(-g)$ and 
it is represented as 
\[\la_0(g)=\half \sqrt{A^2-1}+\half \la_0^{(2)}(0)g^2+\cO(g^3).\]
Let $D_\eps$ be the dilation such that $D_\eps f(x)=f(\eps x)/\sqrt\eps$. 
Thus $D_\eps q D_\eps^{-1}=\eps q$ and 
$D_\eps p D_\eps^{-1}=~p/\eps $, and 
\[D_\eps (p^2+q^2) D_\eps^{-1}=\frac{1}{\eps^2}p^2+\eps^2 q^2.\]
Set $D_{\eps^{1/4}}\gr=\gr(\eps)$. Hence 
we have 
\begin{align*}
&(p^2+\eps q^2)\gr(\eps)=\sqrt\eps \gr(\eps),\\
&(p^2+q^2)\gr(\eps)=(\sqrt\eps+(1-\eps) q^2)\gr(\eps).
\end{align*}
Then 
\begin{align*}
&((\alpha\mp1)p^2+(\alpha\pm1)q^2) \gr(\frac{\alpha\pm1}{\alpha\mp1})=
\sqrt{\alpha^2-1}
\gr(\frac{\alpha\pm1}{\alpha\mp1}). 
\end{align*}
Define the map $\xi$ on $[0,\infty)$ by 
\[\xi(u)=
\pi^{-1/2}\lk e^{-|x|^2 u/2}, 
\lk 1+\lk 1-u\rk q^2\rk\lk p^2+q^2-1\rk ^{-1}
\lk 1+\lk 1-u\rk q^2\rk e^{-|x|^2 u/2}\rk.\]
Note that 
$\xi$ can be expressed differently:
\[\xi(u)=\pi^{-1/2}\lk e^{-|x|^2 /2}, 
\lk 1+\lk \frac{1}{u}-1\rk q^2\rk\lk p^2+\frac{1}{u^2}q^2-\frac{1}{u}\rk ^{-1}
\lk 1+\lk \frac{1}{u}-1\rk q^2\rk e^{-|x|^2 /2}\rk.\]

\begin{lemma}\label{appp}
Recall that $A=\frac{\alpha+\beta}{2}$, $g=\frac{\beta-\alpha}{2}$ 
and 
$A_\pm =A\pm 1$. Then 
we have 
\[\la_0(g)=\half\sqrt{A^2-1}-\half 
\lk\frac{\xi(A_+/A_-)}{A_-}
+\frac{\xi(A_-/A_+)}{A_+}\rk g^2+\cO(g^3).\]
\end{lemma}
\proof 
Let
\[\Phi=\binom{\gr(A_+/A_-)}{\gr(A_-/A_+)}=\binom{\Phi_+}{\Phi_-}.\]
Then 
\[
\half\M{A_-p^2+A_+q^2&0\\0&A_+p^2+A_-q^2}
\Phi=\half\sqrt{A_+A_-}\Phi.\]
Simply we set 
$D_+=D_{(A_+/A_-)^{1/4}}$ and 
$D_-=D_{(A_-/A_+)^{1/4}}$.
Then $D_\pm ^{-1}=D_\mp$ and 
$\Phi_\pm =D_\pm \gr$. 
We have
\begin{align}
\label{m1}
(p^2+q^2)\Phi_\pm=\lk \sqrt{\frac{A_\pm}{A_\mp}}+\lk 1-\frac{A_\pm}{A_\mp}\rk q^2\rk \Phi_\pm.
\end{align}
By formula \eqref{eee} we have
\begin{align*}
&\la_0^{(2)}(0)
=-2\lk
\s_y
h 
\Phi, 
\lk 
Ah 
 +\s_z\half
( a^2+{\add}^2)
 -\half\sqrt{A_+A_-}
\rk ^{-1}
\s_y
h 
\Phi
\rk\\
&=
-4\lk
\binom{-h\Phi_-}{h\Phi_+}, 
\M{
\lk A_-p^2+ A_+q^2 -\sqrt{A_+A_-}\rk^{-1}
&0\\
0&
\lk A_+p^2+ A_-q^2 -\sqrt{A_+A_-}\rk^{-1}
}
\binom{-h\Phi_-}{h\Phi_+}
\rk\\
&=-(\Phi_-, C_-\Phi_-)-(\Phi_+,C_+\Phi_+),
\end{align*}
where $h=\half(p^2+q^2)$ 
and 
\begin{align*}
C_\pm&=
\lk \sqrt{\frac{A_\pm}{A_\mp}}+\lk 1-\frac{A_\pm}{A_\mp}\rk q^2\rk 
\lk A_\pm p^2+ A_\mp q^2 -\sqrt{A_+A_-}\rk^{-1}
\lk \sqrt{\frac{A_\pm}{A_\mp}}+\lk 1-\frac{A_\pm}{A_\mp}q^2\rk\rk\\
&=
\frac{1}{\sqrt{A_+A_-}}\lk \sqrt{\frac{A_\pm}{A_\mp}}+\lk 1-\frac{A_\pm}{A_\mp}\rk q^2\rk 
D_\mp\lk p^2+q^2-1\rk^{-1}D_\mp^{-1}
\lk \sqrt{\frac{A_\pm}{A_\mp}}+\lk 1-\frac{A_\pm}{A_\mp}q^2\rk\rk.
\end{align*}
Here we used 
\eqref{m1}. 
We also see that 
\begin{align*}
&D_+^{-1}
\lk \sqrt{\frac{A_-}{A_+}}+\lk 1-\frac{A_-}{A_+}\rk q^2\rk\Phi_-
=
\sqrt{\frac{A_-}{A_+}}\lk 1+\lk 1-\frac{A_-}{A_+}\rk q^2\rk D_-^2\gr
\end{align*}
Hence 
\begin{align*}
&(\Phi_-,C_-\Phi_-)\\
&=
{\frac{A_-}{A_+}}\frac{1}{\sqrt{A_+A_-}}
\lk
\lk 1+\lk 1-\frac{A_-}{A_+}\rk q^2\rk D_-^2\gr, 
(p^2+q^2-1)^{-1}
\lk 1+\lk 1-\frac{A_-}{A_+}\rk q^2\rk D_-^2\gr\rk\\
&=
{\frac{1}{A_+}}\sqrt{\frac{A_-}{A_+}}\lk
\lk 1+\lk 1-\frac{A_-}{A_+}\rk q^2\rk D_-^2\gr, 
\lk p^2+q^2-1\rk^{-1}
\lk 1+\lk 1-\frac{A_-}{A_+}\rk q^2\rk D_-^2\gr\rk\\
&=\lk D_-^2\gr, M_- D_-^2\gr\rk, 
\end{align*}
and we can also see that 
$(\Phi_+,C_+\Phi_+)
=\lk D_+^2\gr, M_+D_+^2\gr\rk$.
Here 
\[
D_\pm^2\gr(x)=\pi^{-1/4}\exp\lk -\frac{|x|^2}{2}\frac{A_\pm}{A_\mp}\rk \lk \frac{A_\mp}{A_\pm}\rk^{1/4}\]
and 
\begin{align*}
M_\pm=
\frac{1}{A_\mp}\sqrt{\frac{A_\pm}{A_\mp}}
\lk 1+\lk 1-\frac{A_\pm}{A_\mp}\rk q^2\rk 
\lk p^2+q^2-1\rk ^{-1}
\lk 1+\lk 1-\frac{A_\pm}{A_\mp}\rk q^2\rk. 
\end{align*}
Then we obtain that 
$\la_0^{(2)}(0)=-\xi(A_+/A_-)/A_-
-\xi(A_-/A_+)/A_+$ and lemma follows. 
\qed

\begin{theorem}[concavity of eigenvalue curves of $\nec$]
The lowest eigenvalue curve $g\to \la_0(g)$ is concave for sufficiently small $|g|$.
\end{theorem}
\proof
Since 
$p^2+q^2-1$ is a nonnegative self-adjoint operator, 
we see that 
\[\lk \lk 1+\lk 1-u\rk q^2\rk e^{-|x|^2 u/2}, 
\lk p^2+q^2-1\rk ^{-1}
\lk 1+\lk 1-u\rk q^2\rk e^{-|x|^2 u/2}\rk>0\]
for any $u\in\RR$. 
The coefficient of the $g^2$ term of the expansion of 
$\la_0(g)$ with respect to $g$ is 
$-\half (\xi(A_+/A_-)/A_-
+\xi(A_-/A_+)/A_+)<0$ by Lemma \ref{appp}. 
Then 
the theorem follows. 
\qed

\section{Fiber decomposition of NcHO in terms of 2pQRM }
\label{fib}
In this section we decompose $\nec$ into $\rab$'s. 
Let $\rU=e^{-i(\pi/4)\add a}$.
We recall that 
$\rU a^2 \rU^{-1}=ia^2$ and 
$
\rU{\add}^2 \rU^{-1}=-i {\add}^2$ as operator equalities. 
We have
\begin{align}
\label{U}
\nek =(\rs_z\otimes \rU)^{-1} H_{0,\frac{1}{2\sqrt\alpha\beta}}^{\rm 2p}(\rs_z\otimes \rU).
\end{align}
Then by $\nec=\AAA \nek \AAA$, 
we have the proposition below.
\begin{proposition}\label{ab}
Suppose that $\alpha,\beta>0$ and $\alpha\beta>1$. Then 
\begin{align}\label{i}
\nec=\AAA (\rs_z\otimes \rU)^{-1}H_{0,\frac{1}{2\sqrt\alpha\beta}}^{\rm 2p}(\rs_z\otimes \rU) \AAA.
\end{align}
\end{proposition}
In Proposition \ref{ab}, $\rs_z\otimes \rU$ is unitary on $\cH$ but 
unfortunately $\AAA$ is not. 
Hence we can not deduce anything about the spectrum of $\nec$ from 
 that of $H_{0,\frac{1}{2\sqrt\alpha\beta}}^{\rm 2p}$
using Proposition \ref{ab}. 
In this sense Proposition~\ref{ab} is not very useful.

Let $\alpha=\beta$. We see that 
$\sss(Q_{\alpha\alpha}^{\rm NH})=\{\sqrt{\alpha^2-1}(n+\half)\}_{n=0}^\infty$ and $\sqrt{\alpha^2-1}(n+\half)$ is a 
two-fold degenerate eigenvalue for each $n$ by \eqref{qe}. 
Similarly 
by Lemma \ref{re}, 
$\sss(H_{0,g}^{\rm 2p})=\{\sqrt{1-4g^2}(n+\half)\}_{n=0}^\infty$ and 
$\sqrt{1-4g^2}(n+\half)$ is a 
two-fold degenerate eigenvalue for each $n$. 
Hence both $\nec$ and $\rab$ can be seen as extensions of 
the direct sum of the one-dimensional harmonic oscillator. 
Then in special cases such as $\alpha=\beta$ we can immediately identify 
relationships 
between 
$\nec$ and $\rab$.

\begin{lemma}\label{qere}
(1) Suppose that $|g|<\frac{1}{2}$ 
and $\alpha=\beta$ with $\alpha>1$. 
Then 
\[Q_{\alpha\alpha}^{\rm NH}\cong 
\frac{\sqrt{\alpha^2-1}}{\sqrt{1-4g^2}}H_{0,g}^{\rm 2p}.\]
In particular 
$Q_{\alpha\alpha}^{\rm NH}\cong 
\alpha H_{0,\frac{1}{2\alpha}}^{\rm 2p}$. 
(2) 
Suppose that $\alpha=1=\beta$. Then 
\[Q_{11}^{\rm NH}\cong \M{q^2&0\\ 0& p^2}\cong H_{0,\pm\frac{1}{2}}^{\rm 2p}.\]
\end{lemma}
\proof
This follows from \eqref{qe} and 
Lemma \ref{re}. See Figure \ref{F}. 
\qed
\begin{figure}[t]
\[
 \begin{CD}
 Q_{\alpha\alpha}^{\rm NH} @>{\eqref{qe}}>> \half \sqrt{\alpha^2-1}\M{ p^2+ q^2&0\\0& p^2+ q^2} \\
 @VVV @VV{\times \frac{\sqrt{1-4g^2}}{\sqrt{\alpha^2-1}}}V \\
 H_{0,g}^{\rm 2p} @>{Lem. \ref{re}}>> \half \sqrt{1-4g^2}\M{ p^2+ q^2&0\\0& p^2+ q^2}
 \end{CD}
\]
\caption{(1) of Lemma \ref{qere} }
\label{F}
\end{figure}
The purpose of this section is to establish elationships between 
$\nec$ and $\rab$ but for $\triangle>0$ and $\alpha\neq\beta$. 
Both $\rab$ and $\nec$ include similar spectral properties. 
As mentioned in Lemma \ref{qsa}, 
if $\alpha\beta>1$, then $\nec$ is bounded from below and $\sss(\nec)$ is purely discrete. 
However if $\alpha\beta<1$, then 
$\nec$ is unbounded from below. 
Similarly as noted in Lemma~ \ref{ess}, 
if $|g|<\half $, then $\rab$ is bounded from below and $\sss(\rab)$ is purely discrete. 
On the other hand,  if $|g|>\half$, then 
$\rab$ is unbounded from below. 
It is, however, not straightforward to establish relationships between $\nec$ and $\rab$. 
Let
\[c=c_{\alpha\beta}=\M {\sqrt\alpha&0\\0&\sqrt\beta} \rs_z.\]
We can see that $\M {\sqrt\alpha&0\\0&\sqrt\beta} \rs_z=\rs_z\M {\sqrt\alpha&0\\0&\sqrt\beta}$ and
\[\cdd c=c\cdd =\AB,\quad 
(c^{-1})^\ast c^{-1}=
c^{-1}(c^{-1})^\ast= \M{1/\alpha&0\\0&1/\beta}.\]
Define 
\[
\rI_{\alpha\beta}=c\otimes \rU.\]
In \eqref{i} we introduce operators 
$(\rs_z\otimes\rU)\AAA$ and $\AAA(\rs_z\otimes\rU)^{-1}$. 
They can be rewritten as 
$(\rs_z\otimes\rU)\AAA=c\otimes\rU$ and 
$\AAA(\rs_z\otimes\rU)^{-1}=c^\ast \otimes\rU^{-1}$. 
We have
\begin{align*}
&\rI_{\alpha\beta}=c\otimes \rU,\\ 
&\rI_{\alpha\beta}^\ast=c^\ast \otimes \rU^{-1}=c \otimes \rU^{-1},\\
&\rI_{\alpha\beta}^{-1}=c^{-1}\otimes \rU^{-1},\\
&(\rI_{\alpha\beta}^{-1})^\ast=(c^{-1})^\ast \otimes \rU=c^{-1} \otimes \rU.
\end{align*}

\begin{lemma}
\label{hir}
Let $\alpha,\beta>0$ be such that $\alpha\beta>1$, 
and 
$\la\in \CC$. Then 
\begin{align}
\label{e1}
&
(\rI_{\alpha\beta}^{-1})^\ast
(\nec -\la) 
\rI_{\alpha\beta}^{-1}
=\ral -\frac{\alpha+\beta}{2\alpha\beta}\la,\\
\label{e2}
&\nec -\la 
=\rI_{\alpha\beta}^\ast
\left(
\ral -\frac{\alpha+\beta}{2\alpha\beta}\la
\right)
\rI_{\alpha\beta}
\end{align}
hold on $\cD$. 
\end{lemma}
\proof
We have 
\[\nec =c^\ast c\otimes \left( 
\add a +\half\right) + \half\s_y \otimes (-i)(a^2-{\add}^2).\]
Since $(\cdd)^{-1} \s_y c^{-1}=\frac{1}{\sqrt{\alpha\beta}}\s_x$, 
by $\rU\add a\rU^{-1}=\add a$ and 
$\rU(a^2-{\add}^2)\rU^{-1}=i(a^2+{\add}^2)$, 
we see that 
\begin{align*}
(\rI_{\alpha\beta}^{-1})^\ast
(\nec -\la) 
\rI_{\alpha\beta}^{-1}
&=
\one\otimes \left(\add a+\half\right)
+\frac{1}{2\sqrt{\alpha\beta}}
 \s_x \otimes (a^2+{\add}^2)-\la\M{1/\alpha&0\\0&1/\beta} \\
&=
\frac{\alpha-\beta}{2\alpha\beta}\la\s_z\otimes\one+
\one\otimes \left(\add a+\half\right)
+\frac{1}{2\sqrt{\alpha\beta}}
 \s_x \otimes (a^2+{\add}^2)
 - \frac{\alpha+\beta}{2\alpha\beta}\la
 \end{align*}
holds on $\CC^2\otimes C_0^\infty(\RR)$. 
Let $f\in \cD$. 
Since $\CC^2\otimes C_0^\infty(\RR)$ is a core of 
$\ral$, there exists $f_n\in \CC^2\otimes C_0^\infty(\RR)$
so that $f_n\to f$ and $\ral f_n\to \ral f$ as $n\to\infty$.
On the other hand 
$\rI_{\alpha\beta}^{-1} f_n\to
\rI_{\alpha\beta}^{-1} f$ and 
$(\nec-\la)\rI_{\alpha\beta}^{-1} f_n$ converge to 
$\ral f$ as $n\to\infty$, which implies that \eqref{e1} holds true 
on $\cD $ by the closedness of $\nec$. 
\eqref{e2} can be also proven in a similar manner to \eqref{e1}. 
\qed
By Lemma \ref{ess} and the assumption $\alpha\beta>1$, we see that $\sss(\ral)$ is purely discrete. 
We denote the eigenspace of $\ral$ corresponding to eigenvalue $\mu$ by $\cE^{\la}_\mu$. Similarly 
$\cQ_\mu^{\alpha\beta}$ denotes the eigenspace corresponding to  
eigenvalue $\mu$ of $\nec$. 

\begin{lemma}
Let $\la\in \sss(\nec)$. Then 
$\frac{\alpha+\beta}{2\alpha\beta}\la\in \sss(\ral) $. 
\end{lemma}
\proof
Let $f\in\c\rQ_\la^{\alpha\beta}$. Then 
$\rI_{\alpha\beta} f\in \cE^{\la}_{\frac{\alpha+\beta}{2\alpha\beta}\la}$
by Lemma \ref{hir}. 
\qed

Let $\sss(\nec)=\{\la_n\}_{n=0}^\infty$ with $\la_0\leq \la_1\leq\ldots.$ 
In \cite{PW02b} it is shown that 
$\dim \cQ_{\la_n}^{\alpha\beta}\leq 3$ for any $n$. 
From \eqref{ei} it follows that 
$\dim \cQ_{\la_n}^{\alpha\beta}\leq 2$ if $\beta<3\alpha$ or $\alpha<3\beta$. 
If $\alpha=\beta$, 
the lowest eigenvalue $\la_0$ is two-fold degenerate, 
but in \cite{HS13,HS14,wak13} it is proven that 
the lowest eigenvalue $\la_0$ is simple if $\alpha\neq \beta$, i.e., 
\[\dim \cQ_{\la_0}^{\alpha\beta}=\lkk\begin{array}{ll} 1&\alpha\neq \beta,\\
2&\alpha=\beta.
\end{array}
\right.\]
\begin{lemma}\label{mu}
Suppose that $\alpha,\beta>0$ and $\alpha\beta>1$. 
Then $\dim \cQ_{\la}^{\alpha\beta}=\dim 
\cE^{\la}_{\frac{\alpha+\beta}{2\alpha\beta}\la}$ for 
any $\la\in \sss(\nec)$. 
\end{lemma}
\proof
Note that $\ker(\rI_{\alpha\beta}^\ast)=\{0\}$. 
By Lemma \ref{hir} $\rI_{\alpha\beta} \c\rQ_\la^{\alpha\beta}\subset 
\cE^{\la}_{\frac{\alpha+\beta}{2\alpha\beta}\la}$ follows. 
Assume that 
$\cE^{\la}_{\frac{\alpha+\beta}{2\alpha\beta}\la}\setminus \rI_{\alpha\beta} \c\rQ_\la^{\alpha\beta}\neq\{0\}$. 
Let $0\neq g\in 
\cE^{\la}_{\frac{\alpha+\beta}{2\alpha\beta}\la}\setminus \rI_{\alpha\beta} \c\rQ_\la^{\alpha\beta}$. 
Then 
$\rI_{\alpha\beta}^{-1} g\in \c\rQ_\la^{\alpha\beta}$ and hence $g\in \cE^{\la}_{\frac{\alpha+\beta}{2\alpha\beta}\la}$. This is a contradiction. Then 
$\cE^{\la}_{\frac{\alpha+\beta}{2\alpha\beta}\la}=\rI_{\alpha\beta} \c\rQ_\la^{\alpha\beta}$ 
and $\dim \cE^{\la}_{\frac{\alpha+\beta}{2\alpha\beta}\la}\leq \dim \c\rQ_\la^{\alpha\beta}$ follows. 
In a similar manner we can show the opposite inequality. 
Then 
 $\dim \cE^{\la}_{\frac{\alpha+\beta}{2\alpha\beta}\la}=\dim \c\rQ_\la^{\alpha\beta}$ follows. 
\qed
Suppose that $\la\neq\mu$. 
Then in general 
$\cE^{\la}_{\frac{\alpha+\beta}{2\alpha\beta}\la}\not\perp 
\cE^{\mu}_{\frac{\alpha+\beta}{2\alpha\beta}\mu}$. 
We introduce another Hilbert space $\cHa$, on which 
$\cE^{\la}_{\frac{\alpha+\beta}{2\alpha\beta}\la}\perp 
\cE^{\mu}_{\frac{\alpha+\beta}{2\alpha\beta}\mu}$ for $\la\neq \mu$. 
Let $\alpha,\beta>0$ and 
\[\gamma_{\alpha\beta}=\M{1/\alpha&0\\ 0&1/\beta}\otimes \one.\] 
Define an inner product on $\cH$ by
$(f,g)_{\alpha\beta}=(f,\gamma_{\alpha\beta} g)$ 
and the Hilbert space $\cHa$ by 
\begin{align}\label{cha}
\cHa=(\cH, (\cdot,\cdot)_{\alpha\beta}).
\end{align}

\begin{lemma}
Suppose that $\alpha,\beta>0$ and $\alpha\beta>1$. Then (1)-(3) follow. 
\bi
\item[(1)]
$\rI_{\alpha\beta}$ is a unitary operator 
from $\cH$ to $\cH_{\alpha\beta}$. 
\item[(2)] 
Let $\la,\mu\in\sss(\nec)$ with $\la\neq \mu$. 
Then 
$\cE^{\la}_{\frac{\alpha+\beta}{2\alpha\beta}\la}
\perp \cE^{\mu}_{\frac{\alpha+\beta}{2\alpha\beta}\mu}$ in $\cHa$.
\item[(3)] 
 $\rI_{\alpha\beta}\c\rQ_\la^{\alpha\beta}=
\cE^{\la}_{\frac{\alpha+\beta}{2\alpha\beta}\la}$ for any $\la\in \sss(\nec)$ 
and $\cHa=\bigoplus_{\la\in\sss(\nec)} 
\cE^{\la}_{\frac{\alpha+\beta}{2\alpha\beta}\la}$. 
\ei
\end{lemma}
\proof
Since $\rU$ is unitary and $c$ is invertible, 
$\rI_{\alpha\beta}$ is surjective. 
We also have \[(\rI_{\alpha\beta}f, \rI_{\alpha\beta}g)_{\alpha\beta}=
(\rI_{\alpha\beta}f, \gamma_{\alpha\beta}\rI_{\alpha\beta}g)=
(cf, \gamma_{\alpha\beta}cg)=(f,g).\] 
Then (1) follows. 
For any $f\in 
\cE^{\la}_{\frac{\alpha+\beta}{2\alpha\beta}\la}$ 
(resp. $g\in \cE^{\mu}_{\frac{\alpha+\beta}{2\alpha\beta}\mu}$),
$\rI_{\alpha\beta}^{-1} f $ (resp. $\rI_{\alpha\beta}^{-1} g$) 
is 
an eigenvector of $\nec$ corresponding to eigenvalue $\la$ 
(resp. $\mu$) by Lemma \ref{mu}. 
Then
$(\rI_{\alpha\beta}^{-1} f, \rI_{\alpha\beta}^{-1} g)=~0$, but it follows that 
$0=(\rI_{\alpha\beta}^{-1} f, \rI_{\alpha\beta}^{-1} g)
=(f, g)_{\alpha\beta}$. 
Thus (2) follows. 
Let $\la\in\sss(\nec)$. 
Then $\rI_{\alpha\beta}\c\rQ_\la^{\alpha\beta}=
\cE^{\la}_{\frac{\alpha+\beta}{2\alpha\beta}\la}$ follows from Lemma~ \ref{mu}. 
Since 
$\cH=\bigoplus_{\la\in \sss(\nec)}\c\rQ_\la^{\alpha\beta}$, 
we have 
\[\cHa= 
\rI_{\alpha\beta}\cH=\rI_{\alpha\beta}\bigoplus_{\la\in \sss(\nec)}\c\rQ_\la^{\alpha\beta}=
\bigoplus_{\la\in\sss(\nec)} 
\cE^{\la}_{\frac{\alpha+\beta}{2\alpha\beta}\la}.\] 
Then (3) follows. 
\qed
\begin{remark}
(1) The adjoint $\rI_{\alpha\beta}'$ of $\rI_{\alpha\beta}$ as an operator from $\cH$ to $\cHa$ is given by 
\begin{align}\label{ad}
\rI_{\alpha\beta}'=\M{1/\sqrt\alpha&0\\0&1/\sqrt\beta}\rs_z^\ast \otimes \rU^\ast=\rI_{\alpha\beta}^{-1}\neq \rI_{\alpha\beta}^\ast, 
\end{align}
where $\rI_{\alpha\beta}^\ast$ is the adjoint of $\rI_{\alpha\beta}$ as an operator from $\cH$ to $\cH$. 
Thus $(\rI_{\alpha\beta}'f, g)=
(f, \rI_{\alpha\beta} g)_{\alpha\beta}
$. 
(2) We give a remark on \eqref{e1} and \eqref{e2}:
\begin{align*}
&
(\rI_{\alpha\beta}^{-1})^\ast
(\nec -\la) 
\rI_{\alpha\beta}^{-1}
=\ral -\frac{\alpha+\beta}{2\alpha\beta}\la,\\
&\nec -\la 
=\rI_{\alpha\beta}^\ast
\left(
\ral -\frac{\alpha+\beta}{2\alpha\beta}\la
\right)
\rI_{\alpha\beta}. 
\end{align*}
We note that the first equality is analyzed  as a differential equation in \cite{nak24}. 
 Since $\rI_{\alpha\beta}$ is not unitary on $\cH$ but unitary from $\cH$ to $\cHa$, 
above equalities do {\emph not} imply the unitary equivalence between both sides. 
\end{remark}

Let 
$\rP_{\frac{\alpha+\beta}{2\alpha\beta}\la}$ be the projection onto the 
eigenspace $\cE^{\la}_{\frac{\alpha+\beta}{2\alpha\beta}\la}$. 
We set 
\begin{align}
\ral \rP_{\frac{\alpha+\beta}{2\alpha\beta}\la}=
\rar.
\end{align}
Now we are in the position to state the first main result. 
\begin{theorem}[fiber decomposition of $\nec$]
\label{main1}
Let $\alpha,\beta>0$ be such that $\alpha\beta>1$. 
Then 
\[\rI_{\alpha\beta} \nec \rI_{\alpha\beta}^{-1}= \frac{2\alpha\beta}{\alpha+\beta}\bigoplus_{\la\in \sss(\nec)} 
\rar\]
 holds on $\cHa$. Here 
$\bigoplus_{\la\in \sss(\nec)}$ denotes the direct sum in $\cHa$. 
\end{theorem}
\proof
Let $f\in \cE^{\la}_{\frac{\alpha+\beta}{2\alpha\beta}\la}$. 
Then 
$\frac{2\alpha\beta}{\alpha+\beta}\ral f= \la f$ and 
$\nec \rI_{\alpha\beta}^{-1}f=\la \rI_{\alpha\beta}^{-1} f$. 
Let $f,g\in \cE^{\la}_{\frac{\alpha+\beta}{2\alpha\beta}\la}$. Then 
we have 
\begin{align*}
(\rI_{\alpha\beta}^{-1}f, \nec \rI_{\alpha\beta}^{-1}g)_{\alpha\beta}
=
\la (\rI_{\alpha\beta}^{-1}f, \rI_{\alpha\beta}^{-1}g)_{\alpha\beta}
=\la (f, g)=
(f, \ral g).
\end{align*}
On the other hand by \eqref{ad} we see that 
\begin{align*}
(\rI_{\alpha\beta}^{-1}f, \nec \rI_{\alpha\beta}^{-1}g)_{\alpha\beta}
=
(\rI_{\alpha\beta}'f, \nec \rI_{\alpha\beta}^{-1}g)_{\alpha\beta}
=
(f, \rI_{\alpha\beta}\nec \rI_{\alpha\beta}^{-1}g).
\end{align*}
Then it follows that 
\[\rI_{\alpha\beta} \nec\rI_{\alpha\beta}^{-1}= \frac{2\alpha\beta}{\alpha+\beta}\ral \]
on $\cE^{\la}_{\frac{\alpha+\beta}{2\alpha\beta}\la}$. 
Let $\rQ_\la$ be the projection onto $\rQ_\la^{\alpha\beta}$ for each $\la\in \sss(\nec)$. 
Then 
$\rP_{{\frac{\alpha+\beta}{2\alpha\beta}\la}}
 \rI_{\alpha\beta}=
 \rI_{\alpha\beta}\rQ_\la$. 
We see that
\begin{align*}
\frac{2\alpha\beta}{\alpha+\beta}\bigoplus_{\la\in\sss(\nec)}
\ral \rP_{{\frac{\alpha+\beta}{2\alpha\beta}\la}}
&=\bigoplus_{\la\in\sss(\nec)}
\rI_{\alpha\beta} \nec\rI_{\alpha\beta}^{-1}
\rP_{{\frac{\alpha+\beta}{2\alpha\beta}\la}}\\
&=
\rI_{\alpha\beta}\lk
\bigoplus_{\la\in\sss(\nec)}
 \nec Q_{\la}\rk \rI_{\alpha\beta}^{-1}
=\rI_{\alpha\beta}
 \nec \rI_{\alpha\beta}^{-1}.
\end{align*}
Then the proof is completed. 
\qed

\begin{corollary}
Let $\alpha=\beta$ and $\alpha^2>1$. 
Then 
\[\rI_{\alpha\alpha} Q_{\alpha\alpha}^{\rm NH} \rI_{\alpha\alpha}^{-1}= \half \sqrt{\alpha^2-1}
\M{p^2+q^2& 0\\ 0& p^2+ q^2}.\]
\end{corollary}
\proof
Since 
the fiber for $\la$ is $H_{\alpha\alpha}^{\rm 2p}(\la)=H_{0,\frac{1}{2\alpha}}^{\rm 2p}$, 
we have 
\[\rI_{\alpha\alpha} Q_{\alpha\alpha}^{\rm NH} \rI_{\alpha\alpha}^{-1} 
=
\alpha\bigoplus_{\la\in \sss(Q_{\alpha\alpha}^{\rm NH})} 
H_{0,\frac{1}{2\alpha}}^{\rm 2p} 
\rP_{\frac{\la}{\alpha}}
=
\alpha H_{0,\frac{1}{2\alpha}}^{\rm 2p} \bigoplus_{\la\in \sss(Q_{\alpha\alpha}^{\rm NH})} \rP_{\frac{\la}{\alpha}}
=\alpha H_{0,\frac{1}{2\alpha}}^{\rm 2p}\]
and the corollary directly follows. 
\qed

\section{Spectral zeta functions}
\label{3}
In this section we shall construct Feynman-Kac formulas of 
$e^{-t\nec}$ and $e^{-t\rab}$, and investigate the spectral zeta functions. 
\subsection{Probabilistic preparations}
Firstly we consider the spin part. 
Let $\pro N$ be a Poisson process on a probability space 
$(\cY, \cB_\cY, \Pi)$ with the unit intensity.
Let ${\ZZ}_2=\{-1,+1\}$. 
Then 
for $u\in L^2(\ZZ_2)$, 
\[\|u\|^2_{L^2(\ZZ_2)}=\sum_{\s\in\ZZ_2}|u(\s)|^2.\] 
Introducing the norm on $\CC^2$ 
by 
$(u,v)_{\CC^2}=\sum_{i=1}^2 {\bar u_i v_i}$, 
we identify $\CC^2\cong L^2(\ZZ_2)$ 
by $\CC^2\ni u=\vvv{u_1\\u_2}\cong u(\s)$ with 
$u(+1)=u_1$ and $u(-1)=u_2$. 
Note that 
$(u,v)_{\CC^2}= (u,v)_{L^2(\ZZ_2)}$. 
Under this identification 
the Pauli matrices $\s_x,\s_y$ and $\s_z$ are represented as 
the operators $S_x,S_y$ and $S_z$, respectively on $L^2(\ZZ_2)$ by 
$(S_xu)(\s)=u(-\s)$, $(S_yu)(\s)=-i\s u(-\s)$ and 
$(S_zu)(\s)=\s u(\s)$ for $u\in L^2(\ZZ_2)$. 
We define the spin process by 
$S_t=(-1)^{N_t}\s$ for $\s\in{\ZZ}_2$. 
Then it is known that 
\[(u, e^{t\s_x}v)_{\CC^2}=
e^t \sum_{\s\in\ZZ_2}\EE_\Pi^\s [\overline{u(S_0)}v(S_t)].\]
Here $\EE_\Pi^\s\left[\cdots\right] $ denotes the expectation with respect to 
the probability measure $\Pi$ such that $S_0=\s$ a.s. 

Secondly we consider the photon part. 
Let $\rd\mu(x)=\gr^2(x) \rd x$ be a probability measure on $\RR$. 
We define the unitary operator 
$\cU_{\gr}:\LR\to L^2(\RR, \rd \mu)$ by 
\[\cU_{\gr} f=\gr^{-1} f.\]
Let 
$b$ and $\bdd$ be the annihilation operator and the creation operator in $L^2(\RR,\rd \mu)$, which are defined by 
$\gr^{-1}a \gr=b$ and 
$\gr^{-1}\add \gr=\bdd$. 
It is actually given by 
$b=a+\frac{x}{\sqrt2}$ and $\bdd=\add-\frac
{x}{\sqrt2}$. 
They satisfy the canonical commutation relation $[b,\bdd]=\one$, 
and 
\[\bdd+b=\sqrt 2x,\quad \bdd b=-\half \frac{\rd^2}{\rd x^2}+x\frac{\rd}{\rd x}.\]
Let $\pro \rX$ be the Ornstein-Uhrenbeck process on a probability space $(\cX, \cB_\cX, \rP^x)$. 
Here 
${\rP^x}(\rX_0=x)=1$ and $\pro \rX$ satisfies that 
\begin{align*}
\int_{\RR} \EE_{\rP}^x \left[\rX_t \right] \rd \mu(x)=0,\quad 
\int _{\RR} \EE_{\rP}^x \left[\rX_t \rX_s\right] \rd \mu(x) =
\half {e^{-|t-s|}}.
\end{align*}
Here $\EE_{\rP}^x\left[\cdots\right] $ denotes the expectation with respect to 
the probability measure $\rP^x$. 
The generator of $\rX_t$ is given by $-\bdd b$:
\begin{align}\label{F1}
(\phi, e^{-t\bdd b}\psi)_{L^2(\RR,\rd \mu)}=\int_{\RR} \EE_{\rP}^x\left[\overline{\phi(\rX_0)} \psi(\rX_t)\right] \rd \mu(x).
\end{align}
We can compute the density function $\kappa_t$ 
of $\rX_t$ as 
\begin{align}\label{kappa}
\kappa_t(y,x)=\frac{1}{\sqrt{\pi(1-e^{-2t})}}
\exp\left(-\frac{|y-e^{-t}x|^2}{1-e^{-2t}}\right).
\end{align}
Then 
$\EE^x_\rP[f(\rX_t)]=
\int_{\RR} f(y) \kappa_t(y,x) \rd y$.

\subsection{Feynman-Kac formulas for $e^{-t\rab}$ and $e^{-t\nec}$}
In this section we construct a Feynman-Kac formula of $e^{-t\rab}$. 
In \cite{HS24} that of $e^{-t\rabb}$ is studied, which is derived as a special case 
of the formula for 
the  Schr\"odinger operator with spin $\half(p-a)^2-\half \sum_{k=x,y,z}b_k\cdot \s_k+V$. 
See \cite{HIL12}. 
However the Feynman-Kac formula of $e^{-t\rab}$ cannot be constructed 
in the same way as that 
for $e^{-t\rabb}$. 
Set 
\begin{align}
\label{cT}
\rJ= \rs_y\otimes\cU_{\gr}:\cH \to \CC^2\otimes L^2(\RR,\rd \mu),
\end{align} 
where $\rs_y$ is given by \eqref{s2}. 
Then 
\begin{align}
\label{uni}
\rJ\rab \rJ^{-1}= \rak,
\end{align}
where 
\[\rak=-\triangle\s_x\otimes \one+(\one-2g\s_z)\otimes \bdd b +g\s_z\otimes (2x^2-\one)+\half.\]
The operator $\rak$ takes the form:
\[\rak=\M{
(1-2g) \bdd b +g (2x^2-1)+\half&-\tri\\
-\tri& (1+2g) \bdd b -g (2x^2-1)+\half}.\]
Then the off-diagonal part $\MM{0&-\tri\\ -\tri&0}$ of $\rak$ is simple in form, 
and we can construct a Feynman-Kac formula for $e^{-t\rak}$ more easily  than for $e^{-t\rab}$. 
We identify $\cH$ with the set of $L^2$ functions on $\RR\times\ZZ_2$: 
$L^2(\RR\times \ZZ_2,\rd \mu)$. Then for $f,g\in L^2(\RR\times \ZZ_2,\rd \mu)$ we have 
$(f,g)=\sum_{\s\in\ZZ_2}\int _\RR \bar f(x,\s)g(x,\s)\rd\mu(x)$. 
Under this identification we have 
\[
(\rak f)(x,\s)= -\triangle f(x,-\s)+(1-2g\s)(\bdd b f)(x,\s) +g \s (2x^2-1)f(x,\s)+\half f(x,\s).\]
Let 
$$T_s=S_{\triangle s}=(-1)^{N_{\triangle s}}\sigma.$$ 
We define an $(\RR\times\ZZ_2)$-valued stochastic process $\pro q$ on $\cX\times\cY$ 
and $V(\rq_s)$ by 
\begin{align*}
&\rq_t=(\rX_{t(1-2gT_t)},T_t):\cX\times\cY\to \RR\times\ZZ_2,\\
&V(\rq_s)=(2\rX_{s(1-2gT_s)}^2-1)T_s.
\end{align*}
The stochastic process $\pro q$ depends on $\tri$ and $g$. 
When $\tri=0$, $\rq_t=(\rX_{t(1-2g\s)},\s)$, 
and when $g=0$, 
$\rq_t=(\rX_t,T_t)$. 
Let 
\[{\bf E}\left[\ldots\right]=
\half\sum_{\s\in\ZZ_2}\int_{\RR}\EE_{\rP}^x\EE_\Pi^\s\left[\ldots\right] \rd \mu(x). \]
\begin{theorem}[Feynman-Kac formulas]
\label{main2}
Suppose that $|g|<\half$. Then 
we have 
\begin{align*}
&(f, e^{-t\rak}g)_{L^2(\RR\times\ZZ_2,\rd\mu)}=
2e^{\triangle t}e^{-\frac{t}{2}} {\bf E}
\left[
\ov{f(\rq_0)}
g(\rq_t)
e^{-g\int_0^t V(\rq_s)\rd s}
\right],\\
&(f, e^{-t\rab}g)_{L^2(\RR\times\ZZ_2)}=
2e^{\triangle t}e^{-\frac{t}{2}} {\bf E}
\left[
\ov{\rJ f(\rq_0)}
\rJ g(\rq_t)
e^{-g\int_0^t V(\rq_s)\rd s}
\right].
\end{align*}
\end{theorem}
\proof
Let 
\begin{align}
\label{k}
\tilde L=(\one-2g\s_z)\otimes \bdd b +g\s_z\otimes (2x^2-\one)+\half.
\end{align}
Under the identification $\cH\cong L^2(\RR\times \ZZ_2,\rd \mu)$, 
 we can see that 
\[(\tilde Lf)(x,\s)=
(1-2g\s) (\bdd bf)(x,\s) +
g\s (2x^2-1)f(x,\s)+\half f(x,\s),\quad (x,\s)\in\RR\times \ZZ_2.\]
By the Trotter-Kato product formula \cite{tro59, kat78} we see that 
\[e^{-t\rak}=s-\lim_{n\to\infty} \lk e^{\frac{t}{n}\triangle \s_x}e^{-\frac{t}{n}\tilde L}\rk^n.\]
Since we observe that  
\begin{align*}
&(e^{\frac{t}{n}\triangle \s_x}f)(x,\s)=e^{\frac{t}{n}\triangle}\EE_\Pi^\s[f(x,S_{\frac{t}{n}\triangle})],\quad x\in\RR,\\
&(e^{-\frac{t}{n}\tilde L}f)(x,\s)=e^{-\frac{t}{n}\half}\EE_{\rP}^x\left[
e^{-g\int_0^{\frac{t}{n}}(2X^2_{(1-2g\s)s}-1)\s \rd s}f(\rX_{(1-2g\s)\frac{t}{n}},\s)\right],\quad \s\in\ZZ_2,
\end{align*}
together with them we have 
\begin{align*}
(e^{\frac{t}{n}\triangle \s_x}e^{-\frac{t}{n}\tilde L}f)(x,\s)
=
e^{\frac{t}{n}\triangle}e^{-\frac{t}{n}\half}\EE_\Pi^\s\EE_{\rP}^x\left[
e^{-g\int_0^{\frac{t}{n}}(2X^2_{(1-2gS_{\frac{t}{n}\triangle})s}-1)S_{\frac{t}{n}\triangle} \rd s}f(\rX_{(1-2gS_{\frac{t}{n}\triangle})\frac{t}{n}},S_{\frac{t}{n}\triangle})\right]. 
\end{align*}
By the Markov properties of $\pro \rX$ and $\pro N$, we have 
\begin{align*}
&(e^{\frac{t}{n}\triangle \s_x}e^{-\frac{t}{n}\tilde L}e^{\frac{t}{n}\triangle \s_x}e^{-\frac{t}{n}\tilde L}f)(x,\s)\\
&=
e^{\frac{2t}{n}\triangle}e^{-\frac{2t}{n}\half}
\EE_\Pi^\s\EE_{\rP}^x\left[
e^{-g\int_0^{\frac{t}{n}}(2X^2_{(1-2gS_{\frac{t}{n}\triangle})s}-1)S_{\frac{t}{n}\triangle} \rd s}\right.\\
&\hspace{3cm}\times\left. 
\EE_\Pi^{S_{\frac{t}{n}\triangle}}
\EE_{\rP}^{\rX_{(1-2gS_{\frac{t}{n}\triangle})\frac{t}{n}}}\left[
e^{-g\int_0^{\frac{t}{n}}(2X^2_{(1-2gS_{\frac{t}{n}\triangle})s}-1)S_{\frac{t}{n}\triangle} \rd s}f(\rq_{\frac{t}{n}})\right]\right]\\
&=
e^{\frac{2t}{n}\triangle}e^{-\frac{2t}{n}\half}
\EE_\Pi^\s\EE_{\rP}^x\left[
e^{-g\int_0^{\frac{t}{n}}(2X^2_{(1-2gS_{\frac{t}{n}\triangle})s}-1)S_{\frac{t}{n}\triangle} \rd s}
e^{-g\int_0^{\frac{t}{n}}(2X^2_{(1-2gS_{\frac{2t}{n}\triangle})s+(1-2gS_{\frac{2t}{n}\triangle})\frac{t}{n}}-1)S_{\frac{2t}{n}\triangle} \rd s}
f(\rq_{\frac{2t}{n}})\right]\\
&=e^{\frac{2t}{n}\triangle}e^{-\frac{2t}{n}\half}
\EE_\Pi^\s\EE_{\rP}^x\left[
e^{-g\int_0^{\frac{t}{n}}(2X^2_{(1-2gS_{\frac{t}{n}\triangle})s}-1)S_{\frac{t}{n}\triangle} \rd s}
e^{-g\int_{\frac{t}{n}}^{\frac{2t}{n}}
(2X^2_{(1-2gS_{\frac{2t}{n}\triangle})s}-1)S_{\frac{2t}{n}\triangle} \rd s}
f(\rq_{\frac{2t}{n}})
\right].\end{align*}
Repeating these procedures we can see that 
\begin{align*}
\lk\lk e^{\frac{t}{n}\triangle \s_x}e^{-\frac{t}{n}\tilde L}\rk^nf\rk(x,\s)
=e^{t\triangle}e^{-\frac{t}{2}}
\EE_\Pi^\s\EE_{\rP}^x\left[
e^{-g\sum_{j=1}^n 
\int_{\frac{t(j-1)}{n}}^{\frac{tj}{n}}
(2X^2_{(1-2gS_{\frac{tj}{n}\triangle})s}-1)S_{\frac{tj}{n}\triangle} \rd s}
f(\rq_{t})\right].
\end{align*}
 As $n\to\infty$ 
 \[
 \sum_{j=1}^n 
\int_{\frac{t(j-1)}{n}}^{\frac{tj}{n}}
(2X^2_{(1-2gS_{\frac{tj}{n}\triangle})s}-1)S_{\frac{tj}{n}\triangle} \rd s
\to
\int_0^t(2X^2_{(1-2gS_{s})s}-1)S_{s} \rd s\]
almost surely. 
We can see that 
 \begin{align*}
 (g, e^{-t\rak}f)
 &=\lim_{n\to\infty}
 e^{t\triangle}e^{-\frac{t}{2}}
2{\mathbf E}
\left[
\bar g(\rq_0)
e^{-g\sum_{j=1}^n 
\int_{\frac{t(j-1)}{n}}^{\frac{tj}{n}}
(2X^2_{(1-2gS_{\frac{tj}{n}\triangle})s}-1)S_{\frac{tj}{n}\triangle} \rd s}
f(\rq_{t})\right]\\
&=
e^{t\triangle}e^{-\frac{t}{2}}
2{\mathbf E}
\left[
\bar g(\rq_0)
e^{-g
\int_0^t(2X^2_{(1-2gS_{s\triangle})s}-1)S_{s\triangle} \rd s}
f(\rq_{t})\right].
\end{align*}
Then the proof of the first statement is complete. The second statement follows from 
$\rJ\rab \rJ^{-1}= \rak$. 
 \qed
We say that $f\in L^2(\RR\times \ZZ_2,\rd \mu)$ is nonnegative 
if and only if $f\geq0$ almost everywhere. 
We say $f\in L^2(\RR\times \ZZ_2,\rd \mu)$ is positive 
if and only if $f>0$ almost everywhere. 
We denote the set of nonnegative functions by $\cP$, 
and that of positive functions by $\c\rP_+$. 
A bounded operator $T$ on $L^2(\RR\times \ZZ_2,\rd \mu)$ 
is positivity preserving if and only if 
$T \cP\subset \cP$ and 
positivity improving if and only if 
$T \cP\setminus\{0\}\subset \c\rP_+$. 

\begin{corollary}[uniqueness of the ground state]\label{nocrossing}
Suppose that $\triangle\neq0$ and $|g|<\half$. 
Then $e^{-t\rak}$ is positivity improving, and 
the lowest eigenvalue of $\rab$ is simple. 
\end{corollary}
\proof
Assume that $f,g\in \cH$ and $f,g\geq0$. 
Suppose that 
$f(x,\s)>0$ for $(x,\s)\in A\times\{-1\}$ 
and 
$g(x,\s)>0$ for $(x,\s)\in B\times\{-1\}$, 
and the measure of $A\times\{-1\}$ and 
$B\times\{-1\}$ are positive. 
For other cases: 
$f(x,\s)>0$ for $(x,\s)\in A\times\{+1\}$ 
and 
$g(x,\s)>0$ for $(x,\s)\in B\times\{-1\}$ 
or 
$f(x,\s)>0$ for $(x,\s)\in A\times\{+1\}$ 
and 
$g(x,\s)>0$ for $(x,\s)\in B\times\{+1\}$, 
the proof is similar. 
By Theorem \ref{main2} 
we see that 
$0\leq (g, e^{-t\rak}f)$. We shall show that $0\neq (g, e^{-t\rak}f)$. 
We have 
\[(g, e^{-t\rak}f)=2e^{\tri t}e^{-\frac{t}{2}}\sum_{\s\in\ZZ_2}\int _{\RR} g(x,\s)\EE_{\rm P}^x\EE_{\Pi}^\s\left[ f(\rq_t)e^{-g\int_0^t V(\rq_s)\rd s}\right]\rd \mu(x).\]
Assume that 
$(g, e^{-t\rak}f)=0$. Since 
$e^{-g\int_0^t V(\rq_s)\rd s}>0$, 
we have 
\[{\rm supp}g\cap {\rm supp}f(\rq_t)=
{\rm supp}g\cap {\rm supp}f(\rq_t)\cap {\rm supp}e^{-g\int_0^t V(\rq_s)\rd s}\] 
as functions on $\RR\times\ZZ_2\times \cX\times\cY$. 
Then 
\[0=\sum_{\s\in\ZZ_2}\int _{\RR} g(x,\s)\EE_{\rm P}^x\EE_{\Pi}^\s\left[ f(\rq_t)\right]\rd \mu(x).\]
Noticing that 
$\EE_{\Pi}\left[\one_{\{N_{\tri t}={\rm odd}\}}\right]=
\frac{1-e^{-2\tri t}}{2} 
$, 
we see that 
\begin{align*}
0&=
\int _{B} g(x,-1)\EE_{\rm P}^x\EE_{\Pi}^{\s=-1}
\left[f(\rX_{t(1-2g(-1)^{1+N_{\tri t}})}, (-1)^{1+N_{\tri t}})\right]\rd \mu(x)\\
&=
\int _{B} g(x,-1)\EE_{\rm P}^x\EE_{\Pi}^{\s=-1}
\left[\one_{\{N_{\tri t}={\rm odd}\}}
f(\rX_{t(1+2g)}, -1)\right] \gr(x)^2\rd x\\
&=
\int _{B} g(x,-1)\int_{A} 
f(y, -1)
\EE_{\Pi}^{\s=-1}
\left[\one_{\{N_{\tri t}={\rm odd}\}}\right]
\kappa_{t(1+2g)}(y,x) \gr(x)^2\rd x \rd y\\
&=
\frac{1-e^{-2\tri t}}{2} 
\int _{B\times A} 
g(x,-1) 
f(y, -1)
\kappa_{t(1+2g)}(y,x) \gr(x)^2\rd x\rd y.
\end{align*}
Here $\kappa_t$ is the kernel of $\rX_t$ given by \eqref{kappa}. 
Since 
$\int _{B\times A} \rd x \rd y>0$ and 
$g(x,-1)f(y, -1)>0$ for $(x,y)\in B\times A$, 
we have 
\[0=\int _{B\times A} 
g(x,-1) 
f(y, -1)
\kappa_{t(1+2g)}(y,x) \gr(x)^2\rd x\rd y
>0.\]
This is a contradiction. Then 
$0< (g, e^{-t\rak}f)$ for any $f,g\in \cP$ which implies that 
$e^{-t\rak}$ is positivity improving. 
Then the lowest eigenvalue of $\rak$ is simple by the Perron-Frobenius theorem \cite{gro71,far72,FS75}, 
which implies that the lowest eigenvalue of $\rab$ is also simple. 
\qed

Let $e_\tri(g)^{\rm 2p}$ be the lowest eigenvalue of $\rab$. 
By Corollary \ref{nocrossing} 
the eigenvalue curve $g\mapsto e_\tri(g)^{\rm 2p}$ has no crossing to other eigenvalues for $|g|<\half$. 

\begin{corollary}[$\ZZ_4$-symmetry of the ground state]
\label{parity}
Let $\Phi$ be the ground state of $\rab$. Then $\Phi\in \cH_{-1}$. 
\end{corollary}
\proof
By Corollary \ref{nocrossing}, we see that 
$\Phi=\lim_{t\to\infty} e^{-t\rab}\binom{0}{\gr}$ and $P_2\binom{0}{\gr}=-\binom{0}{\gr}$.
Then $P_2\Phi=-\Phi$.
\qed

\begin{remark}
Let $\tri=0$. 
Suppose that $f,g\geq0$ ($f\neq 0$ and $g\neq0$) but 
$f(x,-1)=g(x,+1)=~0$ for any $x\in\RR$. Then 
\[(g, e^{-tL_{0,g}} f)=\sum_{\s\in\ZZ_2}\int _{\RR} g(x,\s)
\EE_{\rm P}^x\EE_{\Pi}^\s\left[ f(\rX_{t(1-2g\s)},\s)\right]\rd \mu(x)=0.\]
Then $e^{-tL_{0,g}}$ is not positivity improving. It is actually seen that 
the lowest eigenvalue of $L_{0,g}$ is two-fold degenerate. 
\end{remark}

We also have another corollary of Theorem \ref{main2}. 
In the case of $\alpha=\beta$ 
it follows that $Q_{\alpha\alpha}^{\rm NH}\cong \half \sqrt{\alpha^2-1}\MM{p^2+q^2&0\\0&p^2+q^2}$. 
Then a Feynman-Kac formula of $e^{-t\nec}$ is trivial 
for $\alpha=\beta$. 
We have a Feynman-Kac type formula of $e^{-t\nec}$ when 
$\alpha\neq \beta$. 
By combining 
$\rI_{\alpha\beta} \nec \rI_{\alpha\beta}^{-1}= \frac{2\alpha\beta}{\alpha+\beta}\bigoplus_{\la\in \sss(\nec)} 
\rar$ and 
the Feynman-Kac formula for $e^{-t\rab}$ 
stated in Theorem \ref{main2}, we can construct a Feynman-Kac formula for $e^{-t\nec}$. 
The $2\times 2$ matrix $\gamma_{\alpha\beta}$ is represented as the function 
\[\gamma_{\alpha\beta}(\s)=\frac{\alpha+\beta}{2\alpha\beta}+\s
\frac{\beta-\alpha}{2\alpha\beta},\quad \s\in\ZZ_2.\]
Then $(u,\gamma_{\alpha\beta} v)_{\CC^2}=\sum_{\s\in\ZZ_2}
\bar u(\s)\gamma_{\alpha\beta}(\s) u(\s)$. 
For each $\la\in \sss(\nec)$ 
we define the stochastic process $\pro {q^{\alpha\beta}}$ by $\pro \rq$ with $\tri=\frac{\alpha-\beta}{2\alpha\beta}\la$ and $g=\frac{1}{2\sqrt{\alpha\beta}}$, i.e., 
\[\rq_t^{\alpha\beta}=
\lk
\rX_{(1-\frac{1}{\sqrt{\alpha\beta}}S_{\frac{\alpha-\beta}{2\alpha\beta}\la t})t},S_{\frac{\alpha-\beta}{2\alpha\beta}\la t} \rk.\]
\begin{corollary}
Let $\alpha,\beta>0$ be such that $\alpha\beta>1$ and $\alpha>\beta$. 
We have 
\begin{align*}
&(f, e^{-t\nec}g)=
2e^{t\frac{\alpha-\beta-\alpha\beta}{\alpha+\beta}} \sum_{\la\in\sss(\nec)}
{\bf E}
\left[
\gamma_{\alpha\beta}
\ov{(\rJ \rI_{\alpha\beta} f_\la)(\rq_0^{\alpha\beta})}
(\rJ\rI_{\alpha\beta} g_\la)(\rq_{\frac{2\alpha\beta}{\alpha+\beta}t}^{\alpha\beta})
e^{
-\frac{1}{2\sqrt{\alpha\beta}}
\int_0^{\frac{2\alpha\beta}{\alpha+\beta}t} 
V(\rq_s^{\alpha\beta})
\rd s}
\right],
\end{align*}
where 
$f_\la$ (resp. $g_\la$) is the projection of $f$ 
(resp.~$g$) to the eigenspace 
$\c\rQ_\la^{\alpha\beta}$. 
\end{corollary}
\proof
Since 
$f=\bigoplus_{\la\in\sss(\nec)}f_\la$ and 
$g=\bigoplus_{\la\in\sss(\nec)}g_\la$, by Theorem \ref{main1}, 
\[(f, e^{-t\nec}g)=
\sum_{\la\in\sss(\nec)}(f_\la, e^{-t\nec}g_\la)=
\sum_{\la\in\sss(\nec)}
(\rI_{\alpha\beta}f_\la, e^{-t\frac{2\alpha\beta }{\alpha+\beta}H_{\alpha,\beta}^{\rm 2p}(\la)} \rI_{\alpha\beta}g_\la)_{\alpha\beta}.
\]
Then the corollary follows from 
Theorem \ref{main2}. 
\qed

\subsection{Spectral zeta function of $\rab$}
Let $\sss(\rab)=\{\mu_n\}_{n=0}^\infty$. 
From a number-theoretical perspective, special attention is given to studying the spectral zeta function, defined as
\[\zeta_{\rm 2p}(s)=\sum_{n=0}^\infty\frac{1}{\mu_n^s}.\]
In \cite{sug18, HS24, RW23} the spectral zeta function of 1pQRM is studied. 
For the 2pQRM we can also consider the spectral zeta functions by path measures, but 
 the statements derived in the present section are similar to \cite{HS24}. 
 Then we only provide outlines of the results. 
The eigenvalues of $\rak$ depend on $g$ and $\tri$. 
Therefore we write $\mu_n=\mu_n(g,\tri)$. 
We have 
\begin{align*}
&\mu_{2m}(0,\tri)=m+\half -\tri,\\
&\mu_{2m+1}(0,\tri)=m+\half +\tri,\\
&\mu_{2m}(g,0)=\mu_{2m+1}(g,0)=\sqrt{1-4g^2}\lk m+\half\rk.
\end{align*}
Thus 
\[\zeta_{\rm 2p}(s)=
\lkk
\begin{array}{ll}
2(1-4g^2)^{s/2}\zeta\lk s;\frac{1}{2}\rk&
\triangle=0,\\
\zeta\lk s;\triangle+\half\rk+\zeta\lk s;-\triangle+\half\rk& g=0,
\end{array}\right.\]
where 
\[\zeta\lk s;\tau\rk=\sum_{n=0}^\infty\frac{1}{(n+\tau)^s}\] denotes the Hurwitz zeta function. 
We establish some technical inequalities 
involving $\rak$, $L_{0,g}$
and $L_{\tri,0}$ to analyze asymptotic behaviors of the spectral zeta function of the 2pQRM. 
We set 
\[\mu_g=\half \sqrt{1-4g^2}.\]
\bl{p1}
Suppose that 
$\mu_g>\triangle$. 
(1) 
Let $0\leq s\leq1$ and $r=n+a$ with $n\in\NN$ and $0<a<1$.
Then 
\begin{align*}
&\|\rak^{-s}\phi\|\leq 
\left(1+\frac{\tri}{\mu_g-\tri}\right)^s\|L_{0,g}^{-s}\phi\|,\\
&\|\rak^{-r}\phi\|
\leq
\left(\frac{1}{\mu_g-\tri}\right)^n
\left(1+\frac{\tri}{\mu_g-\tri}\right)^{a}
\|L_{0,g}^{-a}\phi\|.
\end{align*}
(2) 
Let $0<s\leq 2$ and $r=2n+a$ with $n\in\NN$ and $0<a<2$. 
Then 
\begin{align*}
&(\phi, e^{-t\rak}\phi)\leq 
\frac{1}{t^{s}}
\left(\frac{s}{e}\right)^s
\left(1+\frac{\tri}{\mu_g-\triangle}\right)^s\|L_{0,g}^{-s/2}\phi\|^2,\\
&(\phi, e^{-t\rak}\phi)\leq 
\frac{1}{t^{r}}
\left(\frac{r}{e}\right)^r
\left(\frac{1}{\mu_g-\triangle}\right)^{2n}\left(1+\frac{\triangle}{\mu_g-\triangle}\right)^a\|L_{0,g}^{-a/2}\phi\|^2.
\end{align*}
\el 
\proof
Notice that $\inf\sss(\rak)\geq\mu_g-\triangle$ 
and 
$\|L_{0,g}\phi\|\leq \|\rak\phi\|+\triangle\|\phi\|$. 
We have 
\begin{align}
\|\rak^{-1}\phi\|\leq 
\left(1+\frac{\triangle}{\mu_g-\triangle}\right)\|L_{0,g}^{-1}\phi\|.
\end{align} 
Then L\"owen-Hainz inequality \cite{kat52} we see that 
\[\|\rak^{-s}\phi\|\leq \left(1+\frac{\triangle}{\mu_g-\triangle}\right)^s\|L_{0,g}^{-s}\phi\|\] for any $0\leq s\leq 1$. 
Since $\inf\sss(\rak)\geq\mu_g-\triangle$, 
we also have 
\begin{align*}
\|\rak^{-r}\phi\|
\leq 
\left(\frac{1}{\mu_g-\triangle}\right)^n
\|L_{0,g}^{-a}\phi\|.
\end{align*}
By the first inequality, 
the second one follows. 
The proof of (2) is similar to that of \cite[Lemmas 3.5 and 3.6]{HS24}, 
and we omit it. \qed

\bl{p2}
Suppose that 
$\mu_g>\triangle$.
(1)  
Let $0\leq s\leq1$ and $r=n+a$ with $n\in\NN$ and $0<a<1$.
Then there exist $C_1,C_2>0$ such that 
for $|g|<1/C_1$, 
\begin{align*}
&\|\rak^{-s}\phi\|\leq 
\lkk\frac{1}{1-C_1|g|}\lk 1+\frac{|g|(C_2+\tri)}{\mu_g-\tri}\rk\rkk
^s\|L_{\tri,0}^{-s}\phi\|,\\
&\|\rak^{-r}\phi\|
\leq
\left(\frac{1}{\mu_g-\triangle}\right)^{n}
\lkk\frac{1}{1-C_1|g|}\lk 1+\frac{|g|(C_2+\tri)}{\mu_g-\tri}\rk\rkk^{a}
\|L_{\tri,0}^{-a}\phi\|.
\end{align*}
(2) 
Let $0<s\leq 2$ and $r=2n+a$ with $n\in\NN$ and $0<a<2$. 
Then there exist $C_1,C_2>0$ such that 
for $|g|<1/C_1$, 
\begin{align*}
&(\phi, e^{-t\rak}\phi)\leq 
\frac{1}{t^{s}}
\left(\frac{s}{e}\right)^s
\lkk\frac{1}{1-C_1|g|}\lk 1+\frac{|g|(C_2+\tri)}{\mu_g-\tri}\rk\rkk
^s\|L_{\tri,0}^{-s/2}\phi\|^2,\\
&(\phi, e^{-t\rak}\phi)\leq 
\frac{1}{t^{r}}
\left(\frac{r}{e}\right)^r
\left(\frac{1}{\mu_g-\triangle}\right)^{2n}
\lkk\frac{1}{1-C_1|g|}\lk 1+\frac{|g|(C_2+\tri)}{\mu_g-\tri}\rk\rkk
^a\|L_{\tri,0}^{-a/2}\phi\|^2.
\end{align*}
\el 
\proof
After performing some very tedious calculations considering  $[b,\bdd]=\one$, 
we obtain  that
\[\|(b^2+{\bdd}^2)f\|\leq C_1\|\bdd b f\|+C_2\|f\|.\]
Then we have 
\begin{align*}
&\left\| \M{\bdd b & -\tri\\ -\tri & \bdd b}\phi\right\|\\
&\leq
\left\| \M{\bdd b +g(b^2+{\bdd}^2)& -\tri\\ -\tri & \bdd b-g(b^2+{\bdd}^2)}\phi\right\|
+
\left\| \M{g(b^2+{\bdd}^2)& 0\\ 0 & -g(b^2+{\bdd}^2)}\phi\right\|
\end{align*}
and 
\begin{align*}
\left\| \M{g(b^2+{\bdd}^2)& 0\\ 0 & -g(b^2+{\bdd}^2)}\phi\right\|
&\leq 
|g|C_1\left\| \M{\bdd b &0 \\ 0 &\bdd b}\phi\right\|
+|g|C_2\|\phi\|\\
&\leq 
|g|C_1\left\| \M{\bdd b &-\tri \\ -\tri &\bdd b}\phi\right\|
+|g|(C_2+\tri) \|\phi\|.
\end{align*}
Hence 
$
(1-C_1|g|)\|L_{\tri,0}\phi\|\leq \|\rak\phi\|+|g|(C_2+\tri)\|\phi\|$ 
and we see that 
\begin{align*}
\|L_{\tri,0}\rak^{-1}\phi\|\leq 
\frac{1}{1-C_1|g|}(\|\phi\|+|g|(C_2+\tri)\|\rak^{-1}\phi\|)\leq
\frac{1}{1-C_1|g|}\lk 1+\frac{|g|(C_2+\tri)}{\mu_g-\tri}\rk\|\phi\|. 
\end{align*}
Thus it follows that 
\begin{align*}
\|\rak^{-1}\phi\|\leq 
\frac{1}{1-C_1|g|}\lk 1+\frac{|g|(C_2+\tri)}{\mu_g-\tri}\rk\|L_{\tri,0}^{-1}\phi\|.
\end{align*}
Then proofs of other inequalities are similar to those of Lemma \ref{p1} and 
\cite[Lemmas~3.5 and~3.6]{HS24}, 
and we omit it. \qed

\begin{theorem}\label{main3}
Let $s>1$. 
(1) We have 
$$\lim_{\triangle\to0} \zeta_{\rm 2p}(s)=2(1-4g^2)^{s/2}\zeta\lk s;\frac{1}{2}\rk.$$
(2)
Let $0<\triangle <\half$.  
Then 
$$\lim_{g\to0} \zeta_{\rm 2p}(s)=\zeta\lk s;\triangle+\half\rk+\zeta\lk s;-\triangle+\half\rk.$$
\end{theorem}
\proof
The proof is similar to that of \cite[Theorem 3.7]{HS24}. 
In the proof of (1), we can assume $\mu_g>\triangle$. 
The spectral zeta function $\zeta_{\rm 2p}$ can be presented as 
\begin{align*}
\zeta_{\rm 2p}(s)
&=
\Gamma(s)^{-1}
\int_0^\infty t^{s-1}
{\rm Tr}(e^{-t\rab})\rd t
=
\Gamma(s)^{-1}
\int_0^\infty t^{s-1}
{\rm Tr}(e^{-t\rak})\rd t\\
&=\Gamma(s)^{-1}
\int_0^\infty t^{s-1}
\sum_{n=0}^\infty (f_n, e^{-t\rak}f_n)\rd t
\end{align*}
for any complete orthonormal system $\{f_n\}$ of $L^2(\RR\times \ZZ_2,\rd \mu)$. 
Let $\Phi_{\alpha\,n}\in \cH$, $n\geq0$, $\alpha\in\ZZ_2$, be 
a complete orthonormal system of $\cH$ 
and 
\[L_{0,g}\Phi_{\alpha\,n}=\sqrt{1-4g^2}\lk n+\half\rk \Phi_{\alpha\,n},\quad \alpha\in\ZZ_2.\]
We shall show that 
one can exchange $\lim_{\tri\to0}$ and 
$\int_0^\infty t^{s-1}
\sum_{\alpha\in\ZZ_2}\sum_{n=0}^\infty\ldots\rd t$. 
To show this we construct a function $\rho(t,n)$ independent of $\tri$ 
such that 
$(\Phi_{\alpha\,n}, e^{-t\rak }\Phi_{\alpha\,n}) \leq \rho(t,n)$ and 
$\int_0^\infty t^{s-1} \sum_{\alpha\in\ZZ_2}\sum_{n=0}^\infty \rho(t,n) \rd t<\infty$. 
Set $c_s=2^s\left(\frac{s}{e}\right)^s\geq \left(\frac{s}{e}\right)^s\left(1+\frac{\triangle}
{\mu_g-\triangle}\right)^s$ and 
$a_k=\|L_{\tri,0}^{-k}\Phi_{\alpha\,n}\|^2=\frac{1}{(1-4g^2)^k (n+\half)^{2k}}$ for simplicity. 
Let $1<s\leq2$ and $1<r<s$. 
By Lemma~\ref{p1} we obtain that 
\begin{align*}
(\Phi_{\alpha\,n}, e^{-t\rak }\Phi_{\alpha\,n})
\leq
\frac{c_{r}a_{r/2}}{t^r}
\one_{[0,1)}(t) 
+
\frac{c_{2}a_1}{t^2}\one_{[1,\infty)}(t).
\end{align*}
Set the right-hand side above as $\rho(t,n)$. 
Then 
\begin{align*}
&\int_0^\infty \! 
\sum_{\alpha\in\ZZ_2}
\sum_{n=0}^\infty t^{s-1}
\rho(t,n)\rd t\leq
{\frac{c_{r}}{(1-4g^2)^{r/2}}
\zeta\lk r,\half\rk}
\int_0^1
\! \! t^{s-r-1}
\rd t
+
{\frac{c_{2}}{1-4g^2} 
\zeta\lk 2,\half\rk}
\int_1^\infty
\! t^{s-3}
 \rd t
<\infty.
\end{align*}
Next let $s>2$ and $s<r=2n+a$, where $n\in\NN$ and $0\leq a<2$. 
By Lemma \ref{p1} again we see that 
\begin{align*}
&(\Phi_{\alpha\,n}, e^{-t\rak }\Phi_{\alpha\,n}) \leq
\frac{c_{2}a_1}{t^2}\one_{[0,1)}(t) 
+\frac{a_{a/2}}{t^r}
\left(\frac{r}{e}\right)^r
\left(\frac{1}{\mu_g-\triangle}\right)^{2n}
\left(1+\frac{\triangle}{\mu_g-\triangle}\right)^a
\one_{[1,\infty)}(t).
\end{align*}
Set the right-hand side above as $\rho(t,n)$. 
Then 
\begin{align*}
&\int_0^\infty 
\sum_{\alpha\in\ZZ_2}
\sum_{n=0}^\infty t^{s-1}
\rho(t,n)\rd t\\
&\leq
{\frac{c_2\zeta(2)}{1-4g^2}}
\int_0^1
t^{s-3} 
\rd t+
\left(\frac{r}{e}\right)^r 
\left(\frac{1}{\mu_g-\triangle}\right)^n
\left(1+\frac{\triangle}{\mu_g-\triangle}\right)^a
\frac{\zeta(a)}{(1-4g^2)^{a/2}}\int_1^\infty
\! t^{s-r-1} 
\rd t<\infty.
\end{align*}
Hence by the Lebesgue dominated convergence theorem 
one can exchange 
$\lim_{\tri\to0}$ and $\int_0^\infty t^{s-1} \sum_{\alpha\in\ZZ_2}\sum_{n=0}^\infty\ldots \rd t$
for an $s>1$, 
and 
we have
\begin{align*}
\lim_{\triangle\to0}\zeta_{\rm 2p}(s)=
\Gamma(s)^{-1}
\int_0^\infty t^{s-1}
\sum_{n=0}^\infty \lim_{\triangle\to0} (\Phi_{\alpha n}, e^{-t\rak}\Phi_{\alpha n})\rd t\quad s>1.
\end{align*}
Since 
\begin{align*}
\lim_{\triangle\to0}(f, e^{-t\rak}g)
=
2e^{-\frac{t}{2}} {\bf E}
\left[
\ov{f(\rX_0,\s)}
g(\rX_{t(1-2g\s)},\s)
e^{-g\int_0^t (2\rX_{(1-2g\s)s}^2-1)\s\rd s}
\right]=
(f, e^{-t\tilde L} g), 
\end{align*}
where $\tilde L$ is given by \eqref{k}, 
and $\tilde L\cong L_{0,g}\cong \sqrt{1-4g^2}\MM{\bdd b +\half &0\\0&\bdd b +\half }$, 
we have (1). 
Next we shall show (2). 
Note that $\mu_g>\triangle$ holds true for any $|g|<\half $ if $0<\triangle <\half$. 
In a similar manner but by using Lemma \ref{p2} instead of Lemma \ref{p1}, 
we can also see that 
\begin{align*}
\lim_{g\to0}(f, e^{-t\rak}g)
=
2e^{\triangle t}e^{-\frac{t}{2}} {\bf E}
\left[
\ov{f(\rX_0,T_0)}
g(\rX_{t},T_t)
\right]=
(f, e^{-t L} g). 
\end{align*}
Here 
$L=
\MM{\bdd b+\half+\tri& 0\\ 0 &\bdd b+\half-\tri}$ and 
(2) follows. 
\qed
We show in Theorem \ref{main3} that for any $s>1$
\begin{align*}
&\lim_{\tri\to0}
\sum_{n=0}^\infty
\frac{1}{\mu_n(g,\tri)^s}=\sum_{n=0}^\infty\frac{1}{\mu_n(g,0)^s},\\
&\lim_{g\to0}\sum_{n=0}^\infty\frac{1}{\mu_n(g,\tri)^s}=\sum_{n=0}^\infty\frac{1}
{\mu_n(0,\tri)^s}.
\end{align*}
From this 
the convergences of eigenvalues 
$\mu_n(g,\tri)$ as $\tri\to0$ and $g\to 0$ for each $n\geq0$ can be also shown. 
\begin{corollary}
For each $n\geq0$, we have 
\begin{align*}
&\lim_{\tri\to0}\mu_n(g,\tri)=\mu_n(g,0)=\lkk\begin{array}{ll}\sqrt{1-4g^2}(m+\half)& n=2m,\\
\sqrt{1-4g^2}(m+\half)& n=2m+1,
\end{array}\right.\\
&\lim_{g\to0}\mu_n(g,\tri)=\mu_n(0,\tri)
=\lkk\begin{array}{ll}m+\half-\tri& n=2m,\\
m+\half+\tri& n=2m+1.
\end{array}\right.
\end{align*}
\end{corollary}
\proof 
The proof is similar to \cite[Corollary 3.9]{HS24}. 
Then we omit it. 
\qed
Under a strong condition we can also show an asymptotic behavior of 
the spectral zeta function of the NcHO in Theorem \ref{main5} in Section \ref{7} of Appendix.

\section{1pQRM and one-particle NcHO}
\label{4}
As a counter part of the one-photon quantum Rabi model: 
\[\rabb=\triangle\s_z\otimes \one+\one\otimes 
\lk \add a+\half\rk+ g\s_x\otimes (a+{\add})\]
we define one-particle non-commutative harmonic oscillator $\necc$ by $\nec$ with the interaction $\J\otimes (a^2-{\add}^2)$ replaced by 
$\J\otimes(a-{\add})$: 
\[\necc=\AB\otimes\lk \add a+\half\rk+\half \J \otimes (a-\add)\]
for $\alpha,\beta>0$. 
It can be seen that 
$\rabb$ is self-adjoint on $\cD$ for any $g\in\RR$ and bounded from below, 
$\necc$ is also self-adjoint on $\cD$ and bounded from below for any $\alpha,\beta>0$. 
It is immediate to see that 
\begin{align*}
&Q_{\alpha\alpha}^{\rm 1pNH}\cong \alpha\half \M{p^2+q^2&0\\0&p^2+q^2}-\frac{1}{4\alpha},\\
&H_{0,g}^{\rm 1p}\cong\half \M{p^2+q^2&0\\0&p^2+q^2}-g^2.
\end{align*}
Hence
\[
Q_{\alpha\alpha}^{\rm 1pNH}+\frac{1}{4\alpha}\cong 
\alpha\lk H_{0,g}^{\rm 1p}+g^2\rk.\]

Next we consider the case of $\alpha\neq \beta$ and $\tri\neq0$. 
The analysis for this case is parallel to that of  $\nec$ and $\rab$. 
Thus, we only present an outline.
We define $\tilde \rI_{\alpha\beta}$ by $\rI_{\alpha\beta}$ with 
$\rU$ replaced by $\tilde \rU=e^{-i(\pi/2)\add a}$: 
$\tilde \rI_{\alpha\beta}=c\otimes \tilde\rU$. 
Note that 
$\tilde \rI_{\alpha\beta}:\cH\to\cHa$ is unitary, but not from $\cH$ to $\cH$. 
We see that $\tilde \rU a \tilde \rU^{-1}=ia$ and $\tilde \rU\add \tilde \rU^{-1}=-i\add$. 
In a similar manner to \eqref{e1} and \eqref{e2} we can see that 
\begin{align}
\label{e5}
&
(\tilde \rI_{\alpha\beta}^{-1})^\ast
(\necc -\la) 
\tilde \rI_{\alpha\beta}^{-1}
=\rall -\frac{\alpha+\beta}{2\alpha\beta}\la,\\
\label{e6}
&\necc -\la 
=\tilde \rI_{\alpha\beta}^\ast 
\left(
\rall -\frac{\alpha+\beta}{2\alpha\beta}\la
\right)
\tilde \rI_{\alpha\beta}
\end{align}
 on $\cD$ for each $\la\in \RR$. 
Let 
$\tilde \rP_{\frac{\alpha+\beta}{2\alpha\beta}\la}$ be the projection onto the 
eigenspace $\tilde \cE^{\la}_{\frac{\alpha+\beta}{2\alpha\beta}\la}$ of 
$\rall$. 
We set 
\begin{align}
\rall \tilde \rP_{\frac{\alpha+\beta}{2\alpha\beta}\la}=
\rarr. 
\end{align}

\begin{theorem}[fiber decomposition of $\necc$]
\label{main4}
Let $\alpha,\beta>0$. 
Then 
\[\tilde \rI_{\alpha\beta} \necc \tilde \rI_{\alpha\beta}^{-1}= 
\frac{2\alpha\beta}{\alpha+\beta}\bigoplus_{\la\in \sss(\necc)} 
\rarr \]
holds on $\cHa$. 
Here $\rarr$ is regarded as an operator in $\cHa$. 
\end{theorem}
\proof
The proof is similar to that of Theorem \ref{main1}.
\qed
In a similar manner to \eqref{uni} we have 
$\rJ\rabb \rJ^{-1}= \rakk$,
where 
\[\rakk=-\triangle\s_x\otimes \one+\one \otimes \bdd b +g\sqrt2\s_z\otimes x+\half.\]
In \cite[Lemma 4.1]{HS24} it is shown that 
\begin{align}\label{f}
(f, e^{-t\rabb}g)_{L^2(\RR\times\ZZ_2)}=
2e^{\triangle t}e^{-\frac{t}{2}} {\bf E}
\left[
\ov{\rJ f(\tilde\rq_0)}
\rJ g(\tilde\rq_t)
e^{-g\int_0^t \tilde V(\tilde\rq_s)\rd s}
\right].
\end{align}
Here $\tilde\rq_t=(\rX_t,T_t)$ and $\tilde V(\tilde\rq_s)=\sqrt 2 T_s \rX_s$. 
Let 
$\tilde \rq_t^{\alpha\beta}=
\lk
\rX_t,S_{\frac{\alpha-\beta}{2\alpha\beta}\la t} \rk$. 
Combining Theorem \ref{main4} and \eqref{f} we have the corollary. 
\begin{corollary}
Let $\alpha,\beta>0$ and $\alpha>\beta$. 
We have 
\begin{align*}
(f, e^{-t\necc}g)
=
2e^{t\frac{\alpha-\beta-\alpha\beta}{\alpha+\beta}} 
\!\!\!
\!\!\!
\sum_{\la\in\sss(\necc)}
{\bf E}
\left[
\gamma_{\alpha\beta}
\ov{(\rJ \tilde \rI_{\alpha\beta} f_\la)(\tilde \rq_0^{\alpha\beta})}
(\rJ\tilde \rI_{\alpha\beta} g_\la)(\tilde \rq_{\frac{2\alpha\beta}{\alpha+\beta}t}^{\alpha\beta})
e^{
-\frac{1}{2\sqrt{\alpha\beta}}
\int_0^{\frac{2\alpha\beta}{\alpha+\beta}t} 
\tilde V(\tilde \rq_s^{\alpha\beta})
\rd s}
\right],
\end{align*}
where 
$f_\la$ (resp. $g_\la$) is the projection of $f$ 
(resp.~$g$) to the eigenspace of $\necc$ with eigenvalue~$\la$. 
\end{corollary}

\appendix
\section{Spectrum of $(p+t q)^2+sq^2$}\label{5}
It is known that $p^2+q^2$ is self-adjoint on $\rD(p^2)\cap \rD(q^2)$. We extend this in this section. 
Suppose that $t,s\in\RR$. 
We consider symmetric operators: 
\[T_{t,s}=(p+t q)^2+sq^2.\]
\begin{lemma}
Suppose that $t,s\in\RR$. 
Then $T_{t,s}$ is essentially self-adjoint on any core of $N$. 
\end{lemma}
\proof
Let $N=\half(p^2+q^2+1)\geq1 $. 
We have 
\begin{align*}
&|(f, T_{t,s}g)|\leq C\| N^\frac{1}{2} f\|\| N^\frac{1}{2} g\|,\\
&|( Nf,T_{t,s} g)-(T_{t,s} f, Ng)|
\leq C\| N^\frac{1}{2} f\|\| N^\frac{1}{2} g\|
\end{align*}
for $f,g\in \rD(N)$. 
Then $T_{t,s}$ is essentially self-adjoint on any core of $N$ by the Nelson commutator theorem.
\qed
We denote the self-adjoint extension of $T_{t,s}$ by
 \[\ov{(p+tq)^2+sq^2}=\overline{T_{t,s}\lceil_{\rD(N)}}.\] 

\subsection{Case of $s=0$}
We consider the case of $s=0$. 
Let \begin{align}\label{u}
\cU_t=e^{it q^2/2}
\end{align} be the unitary operator on $\LR$. 
We can see that 
(1) $\cU_t^{-1}p^2\cU_t$ is self-adjoint on $\cU_t^{-1}\rD(p^2)$, 
(2) $\cU_t^{-1} p^2 \cU_t=(p+tq)^2$ holds on $C_0^\infty(\RR)$, 
(3) $C_0^\infty(\RR)$ is a core of both $p^2$ and $(p+tq)^2$. 
\begin{theorem}
\label{s=0}
We have 
\bi
\item[(1)] $\cU_t^{-1} p^2 \cU_t=\ov{(p+tq)^2}$ holds on $\rD(\ov{(p+tq)^2})$, 
\item[(2)] $\rD(\ov{(p+tq)^2})=\cU_t^{-1} \rD(p^2)$,
\item[(3)] $\rD(\ov{(p+tq)^2})= \rD(\ov{(p+uq)^2})$ if and only if $t=u$. 
\ei
\end{theorem}
\proof
By a limiting argument (1) is proven. 
By the uniqueness of self-adjoint extension and 
(1), we see (2). 
Since $\frac{1}{1+x}\in \rD(p^2)$ but 
$\frac{e^{-ivx^2/2}}{1+x}\not\in \rD(p^2)$ for any $v\neq0$, 
we have $\cU_v^{-1}\rD(p^2)\neq \rD(p^2)$ for any $v\neq0$. 
Suppose that 
$\rD(\ov{(p+tq)^2})=\rD(\ov{(p+uq)^2})$. Then 
$\rD(p^2)=\cU_{t-u}^{-1} \rD(p^2)$, but 
since $\cU_v^{-1}\rD(p^2)\neq \rD(p^2)$ for any $v\neq0$, 
$\rD(p^2)=\cU_{t-u}^{-1} \rD(p^2)$ is contradiction. 
Then (3) follows. 
\qed
\begin{example}
$\rD(\ov{(p+q)^2})\neq \rD(\ov{(p-q)^2})$. 
\end{example}

\subsection{Case of $s>0$}
We consider the case of $s>0$. 
\begin{lemma}\label{app4}
Let $s>0$. 
Then 
 $\cU_t$ maps $\rD(p^2+q^2)$ onto itself and
\[\cU_t^{-1}(p^2+sq^2)\cU_t={(p+tq)^2+sq^2}\] on $\rD(p^2+q^2)$. 
\end{lemma}
\proof
Let $h=\half (p^2+ q^2)$. 
Then $\cU_t^{-1}(p^2+sq^2)\cU_t=(p+tq)^2+sq^2$ holds on $C_0^\infty(\RR)$. 
$\cU_t^{-1}(p^2+sq^2)\cU_t$ is self-adjoint on $\cU_t^{-1}\rD(p^2+q^2)$. 
Let $f\in \rD(p^2+q^2)$ and 
$f_n\in C_0^\infty(\RR)$ such that $f_n\to f$ and $hf_n\to hf$ as $n\to\infty$. 
Then $\cU_t^{-1}(p^2+sq^2)\cU_t f_n=\ov{(p+tq)^2+sq^2}f_n$. 
Since $\cU_t f_n\to \cU_t f$ and 
\[\|\ov{(p+tq)^2+sq^2}f_n-\ov{(p+tq)^2+sq^2}f_m\|\leq a\|h(f_n-f_m)\|+b\|f_n-f_m\|\to 0\] as $n,m\to\infty$ with some $a,b\geq0$, 
we see that $\cU_tf\in \rD(p^2+q^2)$ and 
\[\cU_t^{-1}(p^2+sq^2)\cU_tf=
\ov{(p+tq)^2+sq^2} f\] by the closedness of $p^2+sq^2$, but 
$\ov{(p+tq)^2+sq^2}=
{(p+tq)^2+sq^2} $ on $\rD(p^2+q^2)$. 
Then the lemma follows. 
\qed

\begin{theorem}\label{s>0}
Suppose that $t\in\RR$ and $s>0$. Then 
${(p+tq)^2+sq^2}$ is self-adjoint on $\rD(p^2+q^2)$. 
\end{theorem}
\proof
By Lemma \ref{app4}, we have 
$\cU_t^{-1}(p^2+sq^2)\cU_t={(p+tq)^2+sq^2}$ on $\cU_t^{-1}\rD(p^2+q^2)$. 
The left hand side is self-adjoint on $\cU^{-1}\rD(p^2+q^2)$, 
but 
$\cU_t^{-1}\rD(p^2+q^2)=\rD(p^2+q^2)$. 
Then 
the right-hand side is also self-adjoint on $\rD(p^2+q^2)$. 
\qed

\begin{example}
Let $\alpha\beta>1$. Then 
${\left(p\pm \frac{1}{\sqrt{\alpha\beta}}q\right)^2+
\left(1-\frac{1}{\alpha\beta}\right) q^2}$ 
is self-adjoint on $\rD(p^2+q^2)$. 
\end{example}

\subsection{Case of $s<0$}
We consider the case of $s<0$. 
Let $K=\half(pq+qp)$, $L=\half(-p^2+q^2)$ and $N=\half(p^2+q^2)$. 
$N$ is self-adjoint on $\rD(q^2)\cap \rD(p^2)$ and $\sss(N)=\{n+\half\}_{n=0}^\infty$. 
We can see algebraic relations by the canonical commutation relations $[p,q]=-i\one$: 
\begin{align}\label{a}
[N,K]=2i L,\quad 
[L,N]=2i K, \quad
[K,L]=-2i N.
\end{align}
The essential self-adjointness of $K$ is shown in e.g. \cite[Appendix B]{ara83a}. 
We show this for the self-consistency of the paper. 
\begin{lemma}\label{app}
$K$ and $L$ are essentially self-adjoint on any core of $N$ and 
$\sss(K)=\sss(L)=\RR$ and purely absolutely continuous. 
\end{lemma}
\proof
Let $\tilde N=N+\half\geq1$. 
By \eqref{a} we have 
\begin{align*}
&|(f, X g)|\leq C\|\tilde N^\frac{1}{2} f\|\|\tilde N^\frac{1}{2} g\|,\\
&|(\tilde Nf, X g)-(X f, \tilde Ng)|
\leq C\|\tilde N^\frac{1}{2} f\|\|\tilde N^\frac{1}{2} g\|
\end{align*}
for $X=K,L$. 
Then $K$ and $L$ are essentially self-adjoint on any core of $\tilde N$ by 
the Nelson commutator theorem. 
We have $\LR=\LRM\oplus\LRP$. 
Define a unitary operator $\tilde D:\LRP\to \LR$ by 
\[(\tilde Df)(x)={\sqrt 2e^xf(e^{2x})}\] and $D$ by $Df=\widehat {\tilde Df}$, 
where $\hat g$ is the Fourier transform of $g$ and $D$ is introduced in \cite[Appendix B]{ara83a}.
Then $D$ is a unitary operator from $\LRP$ to $\LR$. 
$K$ is reduced to $\LRP$ and $\LRM$. Let 
$K_+=K\lceil_{\LRP}$ and $K_-=K\lceil_{\LRM}$. 
Let $f\in C_0^\infty((-\infty,0))$. 
It is immediate to see that 
$(DK_+f)(k)=k(Df)(k)$ for any $k\in\RR$. 
Since $C_0^\infty((0,\infty))$ is a core of $K$, 
we can extend $(DK_+f)(k)=k(Df)(k)$ for any $f\in \rD(K)\cap \LRP$. 
Hence \[DK_+D^{-1}=k\] holds on $\sss(K)\cap \LRP$, which implies that $\sss(K_+)=\RR$. 
Similarly 
we can also see that $\sss(K_-)=\RR$. Then 
$\sss(K)=\sss(K_-\oplus K_+)=\RR$ and is purely absolutely continuous. 
Since $K=\half (-i) (a^2-{\add}^2)\cong \half(a^2+{\add}^2)=L$ 
by the unitary operator 
$\rU=e^{-i(\pi/4) N}$, 
we also obtain that $\sss(L)=\RR$ and is purely absolutely continuous. 
\qed

\begin{lemma}\label{app5}
Let $s<0$. Then 
Then 
 $\cU_t$ maps $\rD(p^2+q^2)$ onto itself and
$\cU_t^{-1}(p^2+sq^2)\cU_t=\ov{(p+tq)^2+sq^2}$ on $\rD(p^2+q^2)$. 
\end{lemma}
\proof
The proof is similar to that of Lemma \ref{app4}. 
\qed

\begin{theorem}\label{s<0}Suppose that $t\in\RR$ and $s<0$. Then 
$\ov{(p+tq)^2+sq^2}$ is essentially self-adjoint on $\rD(p^2+q^2)$ and 
$\sss(\ov{(p+tq)^2+sq^2})=\RR$. 
\end{theorem}
\proof
The proof is similar to that of Theorem \ref{s>0}. 
\qed

\section{Spectral zeta function of $\nec$}
\label{7}
Let $\sss(\nec)=\{\la_n\}_{n=0}^\infty$. 
Each eigenvalue $\la_n$ depends on $\alpha$ and $\beta$: 
$\la_n=\la_n(\alpha,\beta)$. 
For each $\alpha,\beta>0$ 
the spectral zeta function $\zeta_{\rm NH}$ of $\nec$ is defined by 
\[\zeta_{\rm NH}(s)=\sum_{n=0}^\infty\frac{1}{\la_n^s}.\]
In the case of $\alpha=\beta$, we see that $\la_{2n}=\la_{2n+1}=\sqrt{\alpha^2-1}\lk n+\half\rk$, $n\geq0$,  and $\zeta_{\rm NH}$ is given by 
\[\zeta_{\rm NH}(s)=2(\alpha^2-1)^{-s/2}\zeta\lk s;\half\rk.\]
We consider the asymptotic behavior of 
the spectral zeta function $\zeta_{\rm NH}$ for  $\alpha\neq \beta$ 
as $\beta\to\alpha$.

\begin{theorem}\label{main5}
Suppose that $\alpha$ is sufficiently large. 
Then 
\[\lim_{\beta\to\alpha}
\zeta_{\rm NH}(s)=2({\alpha^2-1})^{s/2}\zeta\lk s;\half\rk.\]
\end{theorem}
\proof
Let $z\in\CC$ with $\Im z\neq 0$. 
Since $\alpha$ is sufficiently large,  
$\left(\add a+\half\right)(Q_{\alpha,\alpha}^{\rm HN}-z)^{-1}$ is bounded. 
Then the operator norm of 
$(\nec-z)^{-1}-(Q_{\alpha,\alpha}^{\rm HN}-z)^{-1}$ can be estimated as 
\begin{align*}
\|(\nec-z)^{-1}-(Q_{\alpha,\alpha}^{\rm HN}-z)^{-1}\|&=
|\beta-\alpha|
\left\|(\nec-z)^{-1}\M{0&0\\0&1}\left(\add a+\half\right) (Q_{\alpha,\alpha}^{\rm HN}-z)^{-1}\right\|\\
&\leq 
\frac{|\beta-\alpha|}{|\Im z|}
\left\|\left(\add a+\half\right) (Q_{\alpha,\alpha}^{\rm HN}-z)^{-1}\right\|\to0
\end{align*}
as $\beta\to\alpha$. 
Then $\lim_{\beta\to\alpha}\la_n(\alpha,\beta)=\la_n(\alpha,\alpha)$ for each~$n$.
Since 
 $(n+\half)\min\{\alpha,\beta\}\sqrt{1-\frac{1}{\alpha\beta}}\leq \la_{2n}\leq\la_{2n+1}$ by Lemma \ref{qsa}, by the Lebesgue dominated convergence theorem 
 we have 
\begin{align*}
\lim_{\beta\to\alpha}
\zeta_{\rm NH}(s)
=
\lim_{\beta\to\alpha}
\sum_{n=0}^\infty \frac{1}{\la_n(\alpha,\beta)^s}
=
\sum_{n=0}^\infty \frac{1}{\la_n(\alpha,\alpha)^s}
=2({\alpha^2-1})^{-s/2}\zeta\lk s;\half\rk.
\end{align*}
\qed

\noindent
{\bf Acknowledgments:} 
We thank Masato Wakayama for appearing \cite{nak24} to us and 
we also thank Daniel Braak for useful comments on the 2pQRM. 
TS was supported by JSPS KAKENHI Grant Numbers 
JP20K20884, JP22H05105, JP23H01077, and JP23K25774, 
and also supported in part by JP20H00119 and JP21H04432.
FH was also financially supported by 
JSPS KAKENHI Grant Numbers  JP20H01808, JP20K20886 and JP23K20217.

From 19th of June 2024 to 10th of July 2024, FH 
was admitted to Kaizuka Hospital in Fukuoka 
due to a ruptured Achilles tendon and coronavirus infection, 
and partially finished this paper during this period of hospitalization. FH would like to express my sincere gratitude to Kaizuka Hospital for providing a quiet environment.

\bibliographystyle{plain}
{ \bibliography{hiro8}}
\end{document}